\documentclass[12pt]{article}
\usepackage{amsmath,amsfonts,amssymb}
\usepackage{graphicx}
\usepackage{subfigure}

\newcommand{\be}{\begin{equation}}
\newcommand{\ee}{\end{equation}}
\newcommand{\ba}{\begin{eqnarray}}
\newcommand{\ea}{\end{eqnarray}}

\usepackage[pdftex,colorlinks=true]{hyperref}
\numberwithin{equation}{section}

\begin{document}
\newcommand{\todo}[1]{{\em \small {#1}}\marginpar{$\Longleftarrow$}}   
\newcommand{\labell}[1]{\label{#1}\qquad_{#1}} 

\rightline{DCPT, KITP}   
\vskip 1cm 

\begin{center} {\Large \bf Boundary conditions for metric fluctuations in Lifshitz}
\end{center} 
\vskip 1cm   

\renewcommand{\thefootnote}{\fnsymbol{footnote}} 
\centerline{Tom\'as Andrade\footnote{tomas.andrade@durham.ac.uk} and Simon F. Ross\footnote{s.f.ross@durham.ac.uk} }
\vskip .5cm 
\centerline{\it Centre for Particle Theory, Department of Mathematical Sciences}
\centerline{\it Durham University, South Road, Durham DH1 3LE, U.K.}

\setcounter{footnote}{0}   
\renewcommand{\thefootnote}{\arabic{footnote}}

\begin{abstract}
We consider the quantisation of linearised fluctuations of the metric and matter fields about a Lifshitz background, exploring the possibility of alternative boundary conditions, allowing the slow fall-off modes to fluctuate. We find that for $z >2$, slow fall-off modes for some of the linearised fluctuations are normalizable, which opens up the possibility of considering alternative boundary conditions. Analysing stability, we find that alternative boundary conditions for the momentum density are allowed, but alternative boundary conditions for the energy density lead to an instability of the type we recently discovered in a similar analysis for scalar fields on a fixed Lifshitz background. Our investigation is in the context of the simple massive vector model, but we would expect the conclusions to be more general. \end{abstract}

\section{Introduction}

The holographic description of field theories with anisotropic scaling symmetry presents an interesting extension of AdS/CFT \cite{Maldacena:1997re,Gubser:1998bc,Witten:1998qj}, which may have valuable applications in condensed matter theory. The simplest example of such a dual description is the Lifshitz metric originally constructed in \cite{Kachru:2008yh}. The geometry is
\begin{equation} \label{lif} 
ds^2 = -r^{2z} dt^2 + r^{2} d\vec{x}^2 + L^2 \frac{dr^2}{r^2},
\end{equation}
where $L^2$ represents the overall curvature scale, and the spacetime has $d+1$ dimensions, so there are $d_s = d-1$ spatial dimensions $\vec{x}$. The asymptotically Lifshitz solutions of a bulk gravity theory are conjectured to provide a dual holographic description of a non-relativistic field theory, with the anisotropic scaling symmetry $t \to \lambda^z t$, $x^i \to \lambda x^i$. This duality is interesting both for its potential application in condensed matter, and as an extension of our understanding of holography and the relation between field theory and spacetime descriptions. The holographic dictionary, relating bulk spacetime  quantities to field theory observables, is now fairly well-developed \cite{Kachru:2008yh,Ross:2009ar,Ross:2011gu}.  As in AdS/CFT, this identifies the leading asymptotic falloff of bulk fields with sources in the dual field theory. 

An interesting early observation in AdS holography \cite{Klebanov:1999tb} was that for some fields, it is possible to introduce an alternative quantisation in the bulk spacetime, where a mode which is subleading in the asymptotic expansion of the field is fixed and the leading piece is allowed to vary. This alternative quantisation leads to a second conformal field theory dual to the same spacetime, with different operator dimensions for the operators dual to the bulk fields whose boundary conditions have been changed. One can also consider mixed boundary conditions, which are dual to renormalisation group flows interpolating between the two conformal field theories, generated by a double-trace deformation of the field theory. For a scalar field, the alternative quantisation is possible when the mass of the field is in the range $m^2_{BF} < m^2 < m^2_{BF} +1$, where $m^2_{BF} = -\frac{d^2}{4}$ is the Breitenlohner-Freedman bound \cite{Breitenlohner:1982jf} for $d$ boundary dimensions. In the bulk spacetime, the restriction to the range $m^2 < m^2_{BF} +1$ comes from the fact that the slow fall-off mode is only normalizable with respect to the usual Klein-Gordon norm for masses in this range. This analysis of alternative boundary conditions in AdS deepened our understanding of the correspondence, through an improved bulk understanding of unitarity and an understanding of the relation of double-trace and more general deformations of the field theory and boundary conditions in the bulk \cite{Witten:2001ua,Berkooz:2002ug}. The early work on scalar fields was subsequently extended to consider vector fields in \cite{Witten:2003ya,Marolf:2006nd}, and for metric perturbations in \cite{Compere:2008us}.  

We recently analysed the possibility of alternative boundary conditions for a scalar field on a fixed Lifshitz background \cite{Andrade:2012xy}, finding that considerations of normalizability would allow alternative boundary conditions in the range $m^2_{BF} < m^2 < m^2_{BF} + z^2$, but that there is a novel type of instability for alternative boundary conditions outside of the range $m^2_{BF} < m^2 < m^2_{BF} +1$. (The normalizability conditions were independently studied in \cite{Keeler:2012mb}). 

In this paper, we address the extension to consider metric fluctuations. We analyse the possibility of alternative boundary conditions for linearised fluctuations of the metric and matter fields, 
about the background \eqref{lif}. Our interest in this calculation is motivated by the hope that it will throw new light on the holographic relation between bulk and boundary if we are able to consider a dynamical boundary geometry. 

This analysis requires some technical preliminaries, so in section \ref{IP}, we first introduce the simple bulk theory we consider and construct the inner product for linearised fluctuations in this theory. In section  \ref{lin}, we review some results of the analysis of the linearised equations of motion of this  theory in \cite{Ross:2009ar}, and extend them by analysing in detail the fluctuations which are independent of the spatial coordinates on the boundary.  

We then investigate the possibility of alternative boundary conditions in section \ref{nbcs}, based on conservation and finiteness of the symplectic product. We find that these are possible for the mode corresponding to the energy density for all $z > 1$, and for the momentum density and the spatial stress tensor for $z >2$. So some of the components of the boundary geometry can be set free; we have a well-defined bulk setup which should be dual to a Lifshitz field theory coupled to a fluctuating geometry. No alternative boundary condition is ever possible for the energy flux; the difference in the behaviour for $z >2$ is perhaps related to its being an irrelevant operator. The fact that alternative boundary conditions are possible for some components of the spatial stress tensor with the usual inner product is a significant qualitative difference from the AdS case. 

In section \ref{spec}, we investigate whether the alternative boundary conditions for the metric fluctuations are subject to instabilities of the kind identified in \cite{Andrade:2012xy}. Because of the complicated coupled nature of the equations, we are only able to conduct numerical investigations. We find that the theory with Neumann boundary conditions for the momentum density appears to be stable, but we find an instability when we allow alternative boundary conditions for the energy density. The salient difference between the two may be that the momentum density is a relevant operator, while the energy density remains marginal. 

In appendix \ref{symm}, we explore the gauge and global symmetries that arise for different boundary conditions. In the appendix \ref{z4}, we briefly discuss the additional issues that arise for $z >4$, where there is a crossover in the modes associated with the momentum density; the mode corresponding to the vev dominates over the mode corresponding to the source for $z >4$. 

\section{Inner product for gravity}
\label{IP}

Following \cite{Taylor:2008tg,Ross:2011gu}, we will study a simple phenomenological theory, 
\begin{equation} \label{Dact}
S =-\frac{1}{16 \pi G_4} \int d^4x \sqrt{-g} \left(R - 2\Lambda -
\frac{1}{4} F_{\mu\nu} F^{\mu\nu} - \frac{1}{2} m^2 A_\mu A^\mu \right) -
\frac{1}{8 \pi G_4} \int d^3 \xi \sqrt{-h} K,
\end{equation}
with $\Lambda = -\frac{1}{2L^2}(z^2+z+4)$, $m^2 L^2=2z$ and
\begin{equation}
A = \alpha r^{z} dt = \sqrt{\frac{2(z-1)}{z}} r^{z} dt.
\end{equation}
Henceforth we set $L=1$, and focus on the case of four bulk spacetime dimensions for simplicity.
 
In this section, we set up the inner product for linearised fluctuations of the metric and massive vector field about any solution of \eqref{Dact}. We then restrict to considering the on-shell linearised fluctuations around the Lifshitz solution \eqref{lif}. 

Following \cite{Lee:1990nz}, we could construct an appropriate conserved current density for the Lagrangian \eqref{Dact} from an appropriate combination of the derivatives of this Lagrangian with respect to the fields and their derivatives. Rather than follow this abstract approach, we will derive the same result by starting from the usual current for linearised gravity
\begin{equation}
j^\mu_g = P^{\mu\nu\alpha\beta\gamma\delta} (h_{2 \alpha \beta}^* \nabla_\nu h_{1 \gamma \delta} - h_{1 \alpha \beta} \nabla_\nu h_{2 \gamma \delta}^*),
\end{equation}
where
\begin{equation}
P^{\mu\nu\alpha\beta\gamma\delta} = \frac{1}{2} (g^{\mu\nu} g^{\gamma (\alpha} g^{\beta) \delta} + g^{\mu (\gamma} g^{\delta) \nu} g^{\alpha \beta} + g^{\mu (\alpha} g^{\beta) \nu} g^{\gamma \delta} - g^{\mu\nu} g^{\alpha \beta} g^{\gamma \delta} - g^{\mu (\gamma} g^{\delta) (\alpha} g^{\beta) \nu} - g^{\mu (\alpha} g^{\beta) (\gamma} g^{\delta) \nu}).
\end{equation}
This current satisfies
\begin{equation}
\nabla_\mu j^\mu_g = h_1^{\alpha \beta} G_{2 \alpha \beta}^{* (1)} - h_2^{* \alpha \beta} G_{1 \alpha \beta}^{(1)}, 
\end{equation}
where $G^{(1)}_{\alpha \beta}$ is the linearised Einstein tensor of the perturbation. Thus, in vacuum Einstein gravity, this is a conserved current, but for \eqref{Dact} the linearised Einstein tensor is related to linearised variations of the massive vector, so this is not separately conserved. From the linearised equations of motion of \cite{Ross:2009ar}, we find that the appropriate piece to add to restore the conservation is
\begin{equation}
j^\mu_a = a_{2 \nu}^* \left(f_1^{\mu \nu} - h_1^{\mu\lambda} F_\lambda^{\ \nu} - h_1^{\beta \nu} F^\mu_{\ \beta} + \frac{1}{2} h_1 F^{\mu \nu} \right) - (1 \leftrightarrow 2),
\end{equation}
where $f_{\mu\nu} = \partial_\mu a_\nu - \partial_\nu a_\mu$ is the linearized field strength, and $F_{\mu\nu}$ is the field strength for the background vector field. Note that this reduces to the usual vector field conserved current $j^\mu_a = a_{2 \nu}^* f_1^{\mu \nu} - a_{1 \nu} f_2^{* \mu\nu}$ when the background vector vanishes. In a background with $A_\mu = 0$, $h_{\mu\nu}$ and $a_\mu$ will be orthogonal with respect to the inner product, but in more general backgrounds they are explicitly coupled.

The current $j^\mu = j^\mu_g + j^\mu_a$ is then conserved when the linearised equations of motion are satisfied, so the inner product 
\begin{equation} \label{mip}
(\{h,a\}_1, \{h,a\}_2) = i \int_\Sigma dx^i \sqrt{h} n^\mu j_\mu
\end{equation}
is independent of the slice $\Sigma$ used to define it, up to possible issues of boundary conditions.

For our Lifshitz spacetime, considering constant time slices, slices of different time will have the same boundary at $r=0$, but will have different intersections with the boundary at infinity, so conservation requires us to check that the boundary conditions at infinity ensure the vanishing of the  flux of $j_\mu$ through this boundary. We will discuss the boundary conditions in section \ref{nbcs} once we have reviewed the asymptotic solution of the equations of motion from \cite{Ross:2009ar}. 

\subsection{Gauge invariance}

The diffeomorphisms which act as gauge symmetries of the spacetime theory should be null directions for the inner product. That is, the inner product should be invariant when we shift one of the linearised solutions (say $h_2^{\mu\nu}, a_{2 \mu}$) by a linearised diffeomorphism which satisfies the boundary conditions.  The way this invariance arises is that the current shifts only by a total derivative when we add such a diffeomorphism. We will evaluate this total derivative explicitly, to verify this invariance. 

We take $h_{2 \alpha \beta} = 2 \nabla_{(\alpha} \xi_{\beta)}$ and $a_{2 \mu} = \xi^\nu \nabla_\nu A_\mu + A_\nu \nabla_\mu \xi^\nu$, and drop the index $1$ on $h_{1 \alpha \beta}$, $a_{1 \mu}$. Then using the background equations of motion and the linearised equations of motion to rewrite expressions in the currents in terms of the background Ricci tensor and the linearised Ricci tensor, we can write the gravitational part of the current as 
\begin{equation}
j^\mu_g = \nabla_\nu X^{\mu\nu}_g - h R^{\mu\nu} \xi_\nu + 2 h^{\mu \beta} R_{\beta \nu} \xi^\nu - 2 \xi_\nu \left( R^{(1)\mu\nu} - \frac{1}{2} g^{\mu\nu} R^{(1)} \right),  
\end{equation}
with 
\begin{equation}
X_{\mu\nu}^g = h \nabla_{[\mu} \xi_{\nu]} + \nabla_{\beta} \xi_{[\mu} h_{\nu]}^\beta + h^\alpha_{[\mu} \nabla_{\nu]} \xi_\alpha + 2 \xi_{[\mu} \nabla_{\nu]} h + 2 \nabla_\beta h^\beta_{[\mu} \xi_{\nu]} + 2 \nabla_{[\mu} h_{\nu]}^\beta \xi_\beta, 
\end{equation}
and the additional part as
\begin{equation}
j^\mu_a = \nabla_\nu X^{\mu\nu}_a + h R^{\mu\nu} \xi_\nu - 2 h^{\mu \beta} R_{\beta \nu} \xi^\nu + 2 \xi_\nu \left(R^{(1)\mu\nu} - \frac{1}{2} g^{\mu\nu} R^{(1)} \right),
\end{equation}
with
\begin{equation}
X^{\mu\nu}_a = A_\alpha \xi^\alpha \left( f^{\mu\nu} - 2h^{\lambda[\mu} F_\lambda^{\ \nu]} + \frac{1}{2} h F^{\mu\nu} \right) + 2 a_\lambda \xi^{[\mu} F^{\nu] \lambda} + a_\lambda \xi^\lambda F^{\mu\nu}.   
\end{equation}
Thus the current is a total derivative for a diffeomorphism, as expected,
\begin{equation}
j^\mu_g + j^\mu_a =  \nabla_\nu (X^{\mu\nu}_g + X^{\mu\nu}_a). 
\end{equation}
Since $X^{\mu\nu}$ is antisymmetric, the integral of $* j$ over a surface then reduces to the integral of $* X$ over the boundary of the surface. 

The boundary term on the boundary at infinity will vanish so long as the inner product is conserved, because we can consider without loss of generality diffeomorphisms with compact support in time and evaluate the inner product on a constant-time surface outside of the support of the diffeomorphism. The boundary term at $r=0$ will be
\begin{equation}
i \int d^{2} x \sqrt{\gamma} n^\mu r^\nu X_{\mu \nu} = i r^{2} \int d^{2} x X_{\hat t \hat r}, 
\end{equation}
where $n^\mu$ is the normal to the surfaces of constant $t$ and $r^\nu$ is the normal to the surfaces of constant $r$ in the surfaces of constant $t$, and the hatted coordinates represent components with respect to an orthonormal frame. If the diffeomorphism $\xi^\mu$ and the perturbation $h^{\mu\nu}, a^\mu$ are smooth as $r \to 0$, then $X_{\hat t \hat r}$ will remain finite, and hence the boundary term will vanish as $r \to 0$. Thus, diffeomorphisms which satisfy the boundary conditions at infinity and are smooth at $r=0$ are null directions for the inner product, and are gauge symmetries of the linearised theory. 

\subsection{Gauge choices}

The form of the inner product can now be simplified by considering an appropriate gauge. At a general level, for an arbitrary solution of \eqref{Dact}, the usual gauge choice to simplify the gravitational current is the De Donder gauge $\nabla^\alpha (h_{\alpha \beta} - \frac{1}{2} g_{\alpha \beta} h) = 0$ used in vacuum gravity. In this gauge,
\begin{equation}
j^\mu_g = \frac{1}{2} \left( h_{2 \alpha \beta}^* \nabla^\mu h_{1 \alpha \beta} - \frac{1}{2} h_2^* \nabla^\mu h_1 \right) - (1 \leftrightarrow 2).
\end{equation}
There is no obvious gauge choice which particularly simplifies $j^\mu_a$, but we can note that the linearised equations of motion imply $\nabla_\nu a^\nu = h^{\lambda \nu}\nabla_\lambda A_\nu$, which we can use to write 
\begin{equation}
j^\mu_a = a_{2 \nu}^* \nabla^\mu a_1^\nu + a_{2 \nu}^* \left( g^{\mu\nu} h_1^{\alpha \beta} \nabla_\alpha A_\beta - h_1^{\mu \lambda} F_\lambda^{\ \nu} - h_1^{\nu\lambda} F^\mu_{\ \lambda} + \frac{1}{2} h_1 F^{\mu\nu} \right) - (1\leftrightarrow 2). 
\end{equation}
If we also separate the metric perturbation into a trace and a traceless part, writing $h_{\alpha \beta} = \tilde{h}_{\alpha \beta} + \frac{1}{4} h g_{\alpha \beta}$ with $\tilde h^\alpha_\alpha = 0$, then
\begin{equation}
j^\mu_a = a_{2 \nu}^* \nabla^\mu a_1^\nu + a_{2 \nu}^* (g^{\mu\nu} \tilde h_1^{\alpha \beta} \nabla_\alpha A_\beta - \tilde h_1^{\mu \lambda} F_\lambda^{\ \nu} - \tilde h_1^{\nu\lambda} F^\mu_{\ \lambda}) - (1\leftrightarrow 2),
\end{equation}
where $\nabla_\nu A^\nu = 0$ has been used in the first term, which follows from the background equations of motion. Thus, the trace of the metric is orthogonal to the other components in this gauge. 

However, this general gauge is not the most convenient choice for the Lifshitz background. Instead, when we analyse the inner product on the Lifshitz background \eqref{lif} we will follow the usual practice of working in a radial  gauge (the analogue of Fefferman-Graham coordinates in the Lifshitz background), where we set $h_{\mu r} = 0$. It is therefore useful to write the inner product explicitly in this gauge before considering the information from the linearised equations of motion. We will at the same time specialise to considering mode solutions in a boundary plane wave basis,
\begin{equation}
h_{\mu\nu} = e^{i \omega t + i k x} h_{\mu\nu}(r), \quad a_\mu = e^{i \omega t + i k x} a_\mu(r),
\end{equation}
choosing spatial coordinates $(x,y)$ such that the mode momentum is aligned along one of the spatial directions. The components $h_{\mu\nu}$ are non-zero only for the indices $\mu, \nu$ along the boundary coordinates $t, x, y$. The integral over the spatial directions in \eqref{mip} will produce a delta function setting $\vec{k_1} = \vec{k_2}$, so we can choose coordinates so that the spatial momenta for both modes are aligned along the same direction. For linearised fluctuations about the Lifshitz spacetime \eqref{lif}, the inner product is then 
\begin{equation}
(\{ h,a \}_1 \{h, a\}_2) = i \int d^{d_s} x dr r^{d_s - 1 -z} j_t,
\end{equation}
where the timelike component of the current is
\begin{eqnarray}
j_t &= & e^{i (\omega_1 - \omega_2) t} \left \{ (\omega_1 + \omega_2) \left[ 2 r^{-2} a_{1x} a_{2x} + 2 r^{-2} a_{1y} a_{2y} + 2r^2 a_{1r} a_{2r} + 2 r^{-4} h_{1xy} h_{2xy} \right. \right. \nonumber \\ &&- \label{modecurr}
\left. r^{-4} (h_{1xx} h_{2yy} + h_{1yy} h_{2xx}) \right] - 2k_1 \left[  r^{-2} (a_{1x} a_{2t} +  a_{1t} a_{2x}) \right. \\ \nonumber  && \left. + r^{-4} (h_{1ty} h_{2xy} + h_{2ty} h_{1xy}) - r^{-4} (h_{1tx} h_{2yy} + h_{2tx} h_{1yy}) \right]- 2i r^2 (a_{1r} a'_{2t} - a'_{1t} a_{2r})  \\ &&- \left. i z \alpha \left[ r^{1-z} (a_{1r} h_{2tt} - a_{2r} h_{1tt}) + r^{z-1} (a_{1r} h_{2xx} + a_{1r} h_{2yy} - a_{2r} h_{1xx} - a_{2r} h_{1yy}) \right] \right \}. \nonumber 
\end{eqnarray}
The inner product will get a divergent contribution from the boundary at infinity if $j_t$ grows faster than $r^{z-2}$. This already indicates that the value $z > 2$ is interesting, as it allows for constant contributions to $j_t$ without producing a divergence.  

\section{Linearised solutions}
\label{lin}

To explicitly evaluate this inner product, we would need solutions of the linearised equations of motion arising from the action \eqref{Dact} for metric and massive vector perturbations around the Lifshitz background \eqref{lif}. The linearised equations of motion are \cite{Ross:2009ar}
\footnote{Note that $h_{\mu\nu}$ denotes
  the perturbation of the metric, and indices are raised and lowered
  with the background metric, so $h^{\mu\nu}$ is the perturbation of
  the metric with the indices raised, not the perturbation of the
  inverse metric. }
\begin{equation}
\nabla_\mu f^{\mu\nu} - \nabla_\mu (h^{\mu\lambda} F_\lambda^{\ \nu})
- \nabla_\mu h^{\beta \nu} F^\mu_{\ \beta}
+ \frac{1}{2} \nabla_\lambda h F^{\lambda \nu} = m^2 a^\nu
\end{equation}
and
\begin{eqnarray}
R_{\mu\nu}^{(1)} &=& \Lambda h_{\mu\nu} + \frac{1}{2} f_{\mu\lambda} F_\nu^{\ \lambda} +
\frac{1}{2} f_{\nu\lambda} F_\mu^{\ \lambda} - \frac{1}{2}
F_{\mu\lambda} F_{\nu\sigma} h^{\lambda \sigma} -\frac{1}{4}
f_{\lambda \rho} F^{\lambda \rho} g_{\mu\nu} + \frac{1}{4} F_{\lambda
  \rho} F_\sigma^{\ \rho} h^{\lambda \sigma} g_{\mu\nu}  \nonumber \\ &&- \frac{1}{8}
F_{\lambda \rho} F^{\lambda \rho} h_{\mu \nu} + \frac{1}{2} m^2
a_\mu A_\nu + \frac{1}{2} m^2 a_\nu A_\mu,
\end{eqnarray}
where $f_{\mu\nu} = \partial_\mu a_\nu - \partial_\nu a_\mu$ and
\begin{equation}
R_{\mu\nu}^{(1)} = \frac{1}{2} g^{\lambda \sigma} [ \nabla_\lambda
  \nabla_\mu h_{\nu \sigma} + \nabla_\lambda \nabla_\nu h_{\mu\sigma} -
  \nabla_\mu \nabla_\nu h_{\lambda \sigma} - \nabla_\lambda
  \nabla_\sigma h_{\mu\nu} ].
\end{equation}

Unfortunately, these cannot be solved in closed form in general. In this section we will show that solutions can be obtained explicitly when the modes do not depend on the spatial coordinates along the boundary, and review results on the asymptotic form of the solutions in general from \cite{Ross:2009ar}. Following \cite{Ross:2009ar,Ross:2011gu}, the slow fall-off parts of the linearised fluctuations will be interpreted as a change in the source for the stress tensor complex, while the fast fall-off parts will give the vevs.

\subsection{Diffeomorphisms and Ward identities}
\label{bdiff}

We study the linearized solutions in the radial gauge defined earlier, where $h_{\mu r}=0$. However this does not wholly fix the diffeomorphism symmetry. The bulk diffeomorphisms which preserve our radial gauge choice are generated by bulk vector fields
\begin{equation} \label{diffeo}
\xi = \sigma(x^\mu) r \partial_r + \phi^\mu(x^\nu) \partial_\mu - \frac{1}{2 z r^{2z}} \partial_t \sigma \partial_t + \frac{1}{2 r^2} \partial_i \sigma \partial_i.
\end{equation}
This corresponds physically to a boundary diffeomorphism $\phi(x^\mu)$ and a boundary Weyl scaling $\sigma(x^\mu)$.\footnote{Note that the diffeomorphism implementing the boundary Weyl scaling is not smooth at $r=0$. This does not imply that the boundary Weyl scaling is not a physical symmetry; it is just a problem with our choice of bulk gauge fixing. For more general choices of bulk gauge there will be smooth diffeomorphisms which realize a boundary Weyl scaling.} These will generically change the slow fall-off parts of the sources, so they change the boundary conditions.\footnote{In appendix \ref{symm}, we discuss the asymptotic isometries, which are the subset of these transformations preserving a given set of boundary data.} 

As a result, the source modes in the linearized solutions can be separated into a diffeomorphism-invariant part and a pure diffeomorphism part. The corresponding feature for the vevs is that they are restricted by a set of holographic Ward identities, 
\begin{equation}\label{ward}
    z {\cal E} - \Pi^i_i = 0, \qquad \partial_t {\cal E} + \partial_i {\cal E}^i = 0, \qquad \partial_t {\cal P}_j + \partial_i \Pi^i_j = 0,
\end{equation}
which are just the conservation equations together with the tracelessness arising from the anisotropic scaling symmetry. In the bulk, these conditions are enforced by the linearised equations of motion in the asymptotic region. The diffeomorphism-invariant modes are conjugate to the remaining free components in the vevs. 

It is useful to separate out the diffeomorphism-invariant part of the source terms from the pure diffeomorphism part. This will clearly be essential for our discussion of stability, where we only want to consider the physical diffeomorphism-invariant modes. But it is also useful to understand what is happening when we change the boundary conditions: passing from Dirichlet to Neumann boundary conditions has a different meaning for the two components. For the diffeomorphism-invariant part, it corresponds to fixing the corresponding vev component instead, but for the pure diffeomorphism component, passing from Dirichlet to Neumann boundary conditions corresponds to passing from treating the diffeomorphism as a global symmetry to treating it as a gauge symmetry. 

\subsection{Zero momentum perturbations}

A special case which enables us to considerably simplify the equations of motion is to consider spatially homogeneous linear perturbations. As we will see, this simplification allows us to solve all the equations of motion in closed form. It will be useful as a subcase for studying the inner product as for time-dependent but spatially homogeneous modes, there is a non-trivial inner product which we can study in detail using the explicit solutions of the linearised equations. In this subsection, we solve the equations of motion for this subsector (this extends the results of \cite{Ross:2009ar}). The inner product for these modes will be discussed in section \ref{ip0}. 

For the zero momentum perturbations, the diffeomorphism degrees of freedom are given by \eqref{diffeo} with $\phi^\mu$ and $\sigma$ now just functions of $t$. The boundary diffeomorphisms $\phi^\mu(t)$ shift the component of $h_{tt}$ which scales as $r^{2z}$ and the components of $h_{ti}$ which scale as $r^2$. These correspond to the sources for the energy density $\mathcal E$ and the momentum density $\mathcal P_i$, so these sources are pure diffeomorphism modes in the zero momentum case. Similarly the boundary Weyl scaling  $\sigma(t)$ shifts the source for the trace of the spatial stress tensor, $\Pi_i^i$. Correspondingly for the sources, the holographic Ward identities \eqref{ward} imply that the vev of the energy density $\mathcal E$ (and hence $\Pi_i^i$) and the momentum density $\mathcal P_i$ must vanish for zero momentum. Thus, the dynamical degrees of freedom for zero momentum are the source and vev for the energy flux $\mathcal E^i$, the traceless part of the spatial stress tensor $\Pi_{ij}^{trf}$ and the scalar operator $\mathcal O_\psi$ coming from the massive vector field. 

To analyse the linearised perturbations, we adopt a parametrization of the components corresponding to that adopted for the constant perturbations in \cite{Ross:2009ar} (since we're setting the momentum to zero, the perturbation preserves the rotational invariance of the Lifshitz background). That is, we set
\begin{equation}
h_{tt} = - r^{2z} f(r)e^{i \omega t}, \quad h_{ti} = (-r^{2z} v_{1i}(r) + r^2 v_{2i}(r))e^{i \omega t} , \quad h_{ij} = ( r^2 k_L(r) \delta_{ij} + r^2 k_{ij}(r))e^{i \omega t} , 
\end{equation}
where $k_{xx} = - k_{yy} = t_d(r)$, $k_{xy} = t_o(r)$, and
\begin{equation}
a_t = \alpha r^z \left( j(r) + \frac{1}{2} f(r) \right)e^{i \omega t} , a_i = \alpha r^z v_{1i}(r)e^{i \omega t} , \quad a_r = i \alpha \frac{p(r)}{r}e^{i \omega t} . 
\end{equation}
The equations of motion then separate into a coupled system for the scalar modes $f, k, j$ and $p$, a coupled system for each $i$ for the vector modes $v_{1i}$ and $v_{2i}$, and decoupled equations for the tensor modes $t_d$ and $t_o$. The pure diffeomorphism modes are the constant parts of $f$, $v_{2i}$ and $k_L$; the other modes are diffeomorphism invariant.

\subsubsection{Zero momentum vector and tensor perturbations}

We consider first the vector and tensor parts, as these have the simplest equations of motion. For the tensor modes, we have 
\begin{equation}\label{eq k=0 for t}
    r^2 t'' + (z+3) r t' + r^{-2 z} \omega^2 t = 0,
\end{equation}
where $t$ is either $t_d$ or $t_o$. It will prove useful to note that \eqref{eq k=0 for t} can be written as a Sturm-Liouville problem with eigenvalue $\omega^2$. The associated Sturm-Liouville inner product will appear in the inner product discussion later. Moreover, we notice that this equation can be simplified by introducing the radial variable $x = r^{-z}$, which gives
\begin{equation}
x z^2 t''- 2 z t' + x \omega^2 t = 0,  
\end{equation}
with primes now denoting derivatives with respect to $x$. The solutions of this equation are 
\begin{equation}
t = t^- t_N + t^+ t_D, 
\end{equation}
where 
\begin{equation}\label{v3D k=0}
    t_D = (2z)^a \Gamma(1 + a) \omega^{-a} x^{a} J_a\left( \frac{\omega x}{z} \right), \quad t_N = (2z)^{-a} \Gamma(1 - a) \omega^{a} x^{a} J_{-a}\left( \frac{\omega x}{z} \right),
\end{equation}
with $a = \frac{z+2}{2z}$. Note that we have normalized the profiles in \eqref{v3D k=0} such that as $r \to \infty$, $t_N \approx 1 + O(r^{-2z})$ and $t_D \approx r^{-z-2} (1 + O(r^{-2z}))$. The interpretation of these coefficients is that $t_{o}^-$, $t_{o}^+$ are the source and vev of the spatial stress tensor component $\Pi_{xy}$, and $t_{d}^-$, $t_{d}^+$ are the source and vev of $\Pi_{xx} - \Pi_{yy}$.  

We should note that neither of the solutions $t_D$ or $t_N$ are regular on their own at $r \to 0$. That is, for pure Neumann or Dirichlet boundary conditions, there are no values of $\omega$ at $k=0$ for which we have regular solutions; this is precisely as we expect, as for conformally invariant boundary conditions regularity at the horizon should give us a dispersion relation of the form $\omega = \alpha k^z$ for some $\alpha$. So the physical problem of zero-momentum modes with Neumann or Dirichlet boundary conditions is trivial. (For mixed boundary conditions there will typically be non-zero values of $\omega$ for which solutions regular at the horizon exist.) However, we will proceed to analyze the inner product in this case as a toy model to develop intuition for the general momentum-dependent case.

For the vector part, we have 
\begin{eqnarray}
\label{vec eom1 k=0}
    v_{1i}'-r^{2-2z} v_{2i}' &=& 0, \\
\label{dec eq v1 k=0}
    r^{-2 z} \omega^2 v_{1i} +  (1+3z) r v_{1i}' + r^2 v_{1i}'' &=& 0. 
\end{eqnarray}
The general solution can be found by first solving \eqref{dec eq v1 k=0} for $v_{1i}$ and then solving \eqref{vec eom1 k=0} for $v_{2i}$. The arbitrary constant mode in $v_{2i}$ is the source for $\mathcal P_i$, which is a pure diffeomorphism as noted above. Introducing the same radial variable $x = r^{-z}$, \eqref{dec eq v1 k=0} reads
\begin{equation}\label{dec eq v1 k=0 rad x}
    x z^2 v_{1i}'' - 2 z^2 v_{1i}' + x \omega^2 v_1  = 0,
\end{equation}
with primes denoting derivatives with respect to $x$. The general solution is
\begin{equation}\label{soln v1 k=0}
    v_{1i} = v_i^- v_{1,N}  + v_i^+ v_{1,D},
\end{equation}
where
\begin{equation}\label{v1 N}
    v_{1,N} = \cos \left( \frac{\omega x}{z} \right) + \frac{\omega x}{z} \sin \left( \frac{\omega x}{z} \right), \quad  v_{1,D} = \frac{3 z^2}{\omega^3} \left[ z \sin \left( \frac{\omega x}{z} \right) - x \omega \cos \left( \frac{\omega x}{z} \right) \right].
\end{equation}
The profiles are normalized such that as $r \to \infty$ we have $v_{1,N} \approx 1 + O(r^{-2z})$ and $v_{1,D} \approx r^{-3z}(1 + O(r^{-2z}))$, corresponding to the slow and fast falloffs. The analysis in \cite{Ross:2009ar} showed that $v_i^-$ is a source for the energy flux, while $v^+_i$ gives its vev. Having obtained the explicit solution for $v_{1i}$, \eqref{vec eom1 k=0} can be solved for $v_{2i}$.\footnote{A closed although rather clumsy expression can be obtained in terms of incomplete Gamma functions. We do not write down the explicit expression since we will not need it to compute the inner products.}  As in the tensor case, there are no non-zero values of $\omega$ for which $v_{1,N}$ or $v_{1,D}$ taken on their own are regular at $r=0$.

\subsubsection{Zero momentum scalar perturbations}

For the scalar perturbations, the equations of motion do not involve $f$ without derivatives, as the constant mode in $f(r)$ corresponds to the source for the energy density, and is a pure diffeomorphism as noted previously. We define $F = f'$.

Taking linear combinations of the equations of motion we find that $F$,$j$,$p$ satisfy the algebraic equations
\begin{eqnarray}
\label{alg F}
    (1+z) r F  &=& 2 (z-1) \left(3 z j + r j' \right) - r^{-2 z} \omega^2 \left( r k_L' + (1+z) k_L\right), \\
\label{alg p}
    2 (z-1) r^{z} p &=&   \omega [ (z-1) k_L - r k_L'], \\
\label{alg j}
  2 (z-1) j &=& r [ (z-4) k_L'-r k_L'' ].
\end{eqnarray}
Solving these algebraic equations we find that the system reduces to a single equation for $S = k_L'$, which reads
\begin{equation}\label{eq S}
    r^2 S'' + r (5+z) S' - ( 2z^2 - 7 z + 1) S = - r^{-2 z} \omega^2 S.
\end{equation}
As in the vector case, we note that \eqref{eq S} can be written as a standard Sturm-Liouville problem with eigenvalue $\omega^2$, which will simplify the computation of the inner products. 
It may also be useful to note that using \eqref{alg j} and \eqref{eq S}, the expression for $F$ can be simplified to
\begin{equation}\label{Fto kL}
    F = -r^{-(1+2z)} \omega^2 k_L + (z-5) k_L'- r k_L''.
\end{equation}
Introducing as before the radial variable $x = r^{-z}$, \eqref{eq S} can be written as
\begin{equation}\label{eq S x}
   x^2 z^2 S'' -4 x z S' - (2z^2 - 7 z + 1 ) S = - x^2 \omega^2 S.
\end{equation}
where the primes in \eqref{eq S x} denote derivatives with respect to $x$. For generic $z$, the general solution of \eqref{eq S x} can be expressed as
\begin{equation}\label{S soln}
    S  = \tilde{A} x^{\frac{4+z}{2 z}} J_{\nu} \left( \frac{x \omega}{z}  \right) +  \tilde{B} x^{\frac{4+z}{2 z}} J_{-\nu} \left( \frac{x \omega}{z}  \right),
\end{equation}
where $\nu = \beta_z/2z $, with $\beta_z^2  = 9z^2 - 20 z + 20$. Having obtained $S$, we can find $k_L$ in terms of generalized hypergeometric functions as
\begin{equation}\label{kL soln}
    k_L = k_L^{-} + s^{+}_\psi x^{2 b_+}  {}_1F_2 \left( b_+ ; b_+ + 1, 1+ \beta_z;  y\right) + s^{-}_\psi x^{2 b_-}  {}_1F_2 \left( b_- ; b_- + 1, 1 -\beta_z; y \right).
\end{equation}
where $b_\pm = (4z)^{-1}( 2+z \pm \beta_z )$, and $y = - x^2 \omega^2/(4 z^2) $. The constant piece $k_L^{-}$ is the source for the trace of the stress tensor $\Pi_{xx} + \Pi_{yy}$, and is a pure diffeomorphism mode as noted previously. We said this diffeomorphism is not smooth; this is reflected in the fact that turning on the constant $k_L^{(0)}$ will produce in \eqref{alg F} a contribution to $f$ which diverges at $r=0$. Thus imposing smoothness at $r=0$ would require us to drop this mode. Again, this is just an issue of a poor choice of gauge. In a more general gauge there will be a smooth diffeomorphism mode which produces a fluctuation with the same asymptotic behaviour as $k_L^{-}$. 

The coefficients $s^{\pm}_\psi$ are related to the vev and source for the scalar operator $\mathcal O_\psi$. They will contribute the only non-trivial part of the inner product in the scalar sector. Note that for $z>2$ the mode parametrized by $s^{-}_\psi$ diverges near the boundary and dominates over the boundary conditions, which implies that it needs to be set to zero. However, we shall find that even in the range $1 <z < 2$ it is non-normalizable. Again, for pure Dirichlet boundary conditions, there are no non-zero values of $\omega$ which give regular solutions at $r=0$. 

\subsection{General perturbations}

For the perturbations with non-zero momentum, we cannot solve the equations in the bulk spacetime in closed form. Here we will just review some of the results of \cite{Ross:2009ar}, writing them in a convenient form and keeping source terms which were dropped in \cite{Ross:2009ar}, as these will be important for our subsequent discussion. 

We use the rotational symmetry to restrict consideration to  mode solutions with the momentum along the $x$ direction. Then the perturbations can be decomposed into scalar and vector modes with respect to the rotation, writing 
\begin{equation}
h_{tt} = - r^{2z} f(r) e^{i\omega t+ i k x}, 
\quad h_{tx} =  [-r^{2z} s_1(r) + r^2 s_2(r)] e^{i\omega t+ i k x} ,
\end{equation}
\begin{equation}
 h_{ty} = [-r^{2z} v_1(r) + r^2 v_2(r)] e^{i\omega t+ i
  k x},
\end{equation}
\begin{equation}
h_{xx} = r^2 \left( k_L(r) + k_T(r) \right)e^{i\omega t+ i k x} , 
  \quad h_{yy} = r^2 \left( k_L(r) -  k_T(r) \right)e^{i\omega t+ i k x},
\end{equation}
\begin{equation}
h_{xy} = r^2 v_3(r) e^{i\omega t+ i k x},
\end{equation}
and
\begin{equation}
a = \alpha r^z e^{i\omega t+ i k x} \left [ \left(j(r) + \frac{1}{2} f(r) \right) dt +  s_1(r) dx  + v_1(r) dy + i\frac{p(r)}{r^{z+1}} dr \right].
\end{equation}
The functions $v_1, v_2, v_3$ represent divergence-free vector
excitations, while the other functions are scalars or scalar-derived
vectors with respect to the rotational symmetry in the $x,y$
plane. The scalar modes and vector modes decouple, so we can analyse
them separately.

{\it Note} that the notation here is slightly different from in \cite{Ross:2009ar}; we have dropped some factors of $k$ which were included in the definitions of the functions there for convenience in solving the equations of motion. For interpreting the form of expressions appearing in the inner product, it is better to have a notation where the functions can be directly interpreted as corresponding to components of the metric and massive vector field, without any factors of $k$ or $\omega$ intervening. 

\subsubsection{Vector perturbations}
\label{vp}

In this notation, the linearised equations of motion in the vector sector are \cite{Ross:2009ar} 
\begin{eqnarray}
\label{vec eom1}
  \omega (r v_1' - r^{-2(z-1)} r v_2') &=& - k r v_3' \\
\label{vec eom2}
 r^2 v_1'' + (2 z + 1) r v_1' + z r^{-2(z-1)} r v_2' &=& \left ( \frac{k^2}{r^2} - \frac{\omega^2}{r^{2z}} \right) v_1 \\
\label{vec eom3}
    r^2 v_3'' + (z + 3) r v_3' + \frac{k \omega v_1}{r^2} - \frac{k \omega v_2}{r^{2z}} &=& - \frac{\omega^2}{r^{2z}} v_3.
\end{eqnarray}
The asymptotic form of the solution is 
\begin{eqnarray} \label{v1a}
v_1 &=& v_1^- + \frac{z}{2(z-1)(z+2)} \frac{k c_y}{r^{z+2}} + v_1^+ r^{-3z} + \ldots, \\
v_2 &=& v_2^- + \frac{z-2}{2(z-1)(z-4)} k c_y r^{z-4} + \frac{3z}{z+2} v_1^+ r^{-z-2} + \ldots, \\ 
 v_3 &=& v_3^- - \frac{1}{z+2} \omega c_y r^{-z-2} + \frac{(z-1)}{(3z^2 + 8z+4)} k \omega v_1^+ r^{-3z-2} + \ldots.  \label{v3a}
\end{eqnarray}
We have only written the leading source and vev terms here. There are subleading corrections involving the derivatives of the leading sources,  some of which appear before the leading vev terms. Note that there is a mixing between the different modes, so subleading terms involving one of the sources can appear in the other functions. The explicit form of these subleading terms is not illuminating, so we will not write it explicitly. We need the full form as an asymptotic expansion in the numerical investigation of stability, but for our other calculations the terms written here are sufficient; for the inner product we just need to check that the term coming from the inner product between the leading source terms does not produce a divergence at large $r$, and in the flux calculation we are mostly interested in the finite part which comes from the interaction of the leading source and vev terms.  

The interpretation of the different modes here is that $v_1^-$ and $v_1^+$ give the source coupling to $\mathcal E_y$ and its vev. For $z >2$ we know $\mathcal E_y$ is an irrelevant operator, so we need to set $v_1^- =0$.  Also $v_2^-$ is the source coupling to $\mathcal P_y$ and $v_3^-$ is the source coupling to $\Pi_{xy}$, but their vevs are related, 
\begin{equation}
\langle \mathcal{P}_y \rangle = kc_y, \quad  \langle \Pi_{xy} \rangle = - \omega c_y.
\end{equation}
The two vevs are determined by a single bulk mode because they are related by a Ward identity.  The corresponding diffeomorphism is the vector part of \eqref{diffeo}, $\xi^\mu = \phi^y(t,x) \partial_y$, which shifts $v_3^-$ and $v_2^-$. The diffeomorphism invariant combination is $\partial_t v_3^- - \partial_x v_2^-$.  We will see later when we consider the fluxes that we can take a boundary condition where we fix the sources $v_2^-, v_3^-$ or a boundary condition where we fix the vev $c_y$, and treat the diffeomorphism as a gauge symmetry. In the stability analysis, with the latter boundary condition we will work in a gauge where we choose the diffeomorphism such that the source for the spatial stress tensor component remains fixed. The choice of boundary conditions for the diffeomorphism-invariant modes will then correspond to the choice of either Dirichlet or Neumann boundary conditions for the momentum density.

\subsubsection{Scalar perturbations}
\label{sp}

In the scalar sector, the function $p({}r)$ is determined algebraically in terms of the other functions, which satisfy a system of three second order and three first order equations of motion \cite{Ross:2009ar}: 
\begin{eqnarray}\label{eq s 0}
\nonumber
	\frac{k^2}{r^2} f + (1+z) r f' +6 z (1-z) j + 2 (1-z) r j' + \left( \frac{k^2}{r^2} - (1+z) \frac{\omega^2}{r^{2z}}   \right) k_L 
	 \\ 
\nonumber	
	+ \frac{\omega^2}{r^{2z}} r k_L' - \frac{k^2 }{r^2} k_T   - 2 (1+z) r k_T'  - \frac{2 k \omega}{r^2} s_1  + \frac{(k^2 - 4r^2 (1+z)) \omega}{2k r } s_1'
	\\ 
     +  (1+z) \frac{k \omega}{r^{2z}} s_2  - \frac{(k^2 - 4 r^2(1+z))}{2r} \frac{\omega}{kr^{2z}} s_2' = 0 
\end{eqnarray}
\begin{eqnarray}
\nonumber
	\frac{2 k^2}{r^2} f + (2z - 3) r f' - r^2 f'' - 8 (z-1)^2 j + \left( \frac{k^2}{r^2} - 4 \frac{\omega^2}{r^{2z}} \right) k_L  \\ 
\nonumber	
	- \frac{k^2}{r^2} k_T  - 2 (2z+1) r k'_T - \frac{4 k \omega}{r^2} s_1 - 2 (2z+1) \frac{\omega}{k} r s_1' \\
	 +  4 k \omega r^{-2z} s_2   + 2 (2z+1) r^3 \frac{\omega}{k r^{2z}} s_2'  = 0 
\end{eqnarray}
\begin{equation}\label{eq s 2}
	r f' + 2(1-z) j + + r k_L' - r k_T' - r \frac{\omega}{k} s_1' + \frac{\omega}{k r^{2z}} r^3 s_2' = 0 
\end{equation}
\begin{equation}
	\frac{k^2 f}{2 r^2}   -  \frac{\omega^2}{r^{2z}} k_T -  (z+3) r k_T' -  r^2 k_T''  - \frac{k \omega}{r^2} s_1 +  \frac{k \omega}{r^{2z}} s_2= 0
\end{equation}
\begin{eqnarray}
\nonumber
	k \omega (- 2 r f' + 4 (z-1) j + 2 (2-z) k_L - 2 k^2 k_T   + 2 k^2 r k_T' ) \\
\nonumber
	+ [k^2 r^{2z-1} + 2r( (3+z) r^{2z} + \omega^2 )] s_1' + 2 r^{2(z+1)} s_1'' \\
	 + 2 k^2 (z-1) s_2 - \left[ k^2  + 2 r^2 \left( 5 - z + \frac{\omega^2}{r^{2z}} \right) \right] r s_2'   
	 - 2 r^4 s_2 ''  = 0 
\end{eqnarray}
\begin{eqnarray}\label{eq s 5}
\nonumber
	k \omega (2 k^2 f - 2 k^2 r f' + 4 k^2 (z-1) j - 2( k^2(z-2) + 4 r^{2-2z} \omega^2  ) k_L - 2 k^2 k_T )\\
	 \nonumber  + 2 r k \omega (k^2 - 2 r^2) k_T' 
	- 4 k^2 \omega^2 s_1 + [  k^4 r^{2z-1} - 4 r^3 \omega^2 - 2 k^2 r (  r^{2z} (z-2)  - \omega^2 )  ] s_1'     \\
+ [ 2 k^4 (z-1) + 4 k^2 r^{2-2z} \omega^2  ] s_2	+ [ 4 r^{5-2z} \omega^2 - k^4 r - 2 k^2 r^{3-2z} (r^{2z} z + \omega^2)  ] s_2 ' = 0
\end{eqnarray}
 The asymptotic solution of these equations is
\begin{eqnarray} \label{fa} 
f &=& f^- - \frac{2(5z-2+\beta_z)}{(z+2-\beta_z)} \frac{s^-_\psi}{r^{\frac{1}{2} (z+2-\beta_z)}} + \frac{z}{z^2-4} \frac{k c_t}{r^{z+2}} - \frac{2(5z-2-\beta_z)}{(z+2+\beta_z)} \frac{s^+_\psi}{r^{\frac{1}{2} (z+2+\beta_z)}} + \ldots,\\
j &=& \frac{(z+1)}{(z-1)} \frac{s^-_\psi}{r^{\frac{1}{2} (z+2-\beta_z)}}- \frac{z(z+1)}{4(z-2)(z-1)} \frac{k c_t}{r^{z+2}} + \frac{z+1}{z-1} \frac{s^+_\psi}{r^{\frac{1}{2} (z+2+\beta_z)}} + \ldots, \\
 k_T &=& k_T^- - \frac{z}{2(z+2)} \frac{kc_t}{r^{z+2}} - \frac{1}{(z+2)} \frac{\omega c_x}{r^{z+2}} + \ldots, \\
k_L &=& k_L^- + \frac{2(3z-4+\beta_z)}{(z+2-\beta_z)} \frac{s^-_\psi}{r^{\frac{1}{2} (z+2-\beta_z)}} + \frac{z}{2(z^2-4)} \frac{k c_t}{r^{z+2}} + \frac{2(3z-4-\beta_z)}{(z+2+\beta_z)} \frac{s^+_\psi}{r^{\frac{1}{2} (z+2+\beta_z)}} + \ldots, \\
 s_1 &=& s_1^-  + \frac{1}{6(z-1)} \frac{\omega c_t}{r^{3z}} + \frac{z}{2(z-1)(z+2)} \frac{kc_x}{r^{z+2}} + \ldots, \\
 s_2 &=& s_2^-  + \frac{z}{2(z-1)(z+2)} \frac{\omega c_t}{r^{z+2}} + \frac{(z-2)}{2(z-1)(z-4)} k c_x r^{z-4} + \ldots . \label{s2a}
\end{eqnarray}
As for the vector modes above, we are not writing the full asymptotic expansion explicitly, but only retaining a set of terms we are interested in for calculations below. There are subleading corrections involving the derivatives of the leading sources,  with a mixing between the different modes, so subleading terms involving one of the sources can appear in the other functions. 

There are nine free modes in the asymptotic solution: The interpretation of these modes is that $f^-$ is the source for $\mathcal E$, $k_T^-$ is the source for $\Pi_{xx} - \Pi_{yy}$, $k_L^-$ is the source for $\Pi_{xx} + \Pi_{yy}$, $s_1^-$ is the source for $\mathcal E_x$, and $s_2^-$ is the source for $\mathcal P_x$. As in the vector sector, we should set $s_1^- = 0$ for $z >2$. The vevs are given by 
\begin{equation}
\langle \mathcal{E} \rangle = kc_t, \quad \langle \mathcal{E}_x \rangle = -\omega c_t,
\end{equation}
\begin{equation}
\langle \mathcal{P}_x \rangle = kc_x, \quad  \langle \Pi_{xx} \rangle = - \omega c_x, \quad \langle \Pi_{yy} \rangle = z k c_t + \omega c_x. 
\end{equation}
The modes $s^\pm_\psi$ give the vev and the source for the scalar operator $\mathcal O_\psi$. As in the zero momentum case, we must also set $s^-_\psi = 0$ for $z >2$. 

The diffeomorphism here is the scalar part of \eqref{diffeo}, which is everything except the $\phi^y$ term, where now $\phi^\mu$, $\sigma$ are functions of $t$ and $x$. The component $\phi^t$ shifts $f^-$ and $s_1^-$, the component $\phi^x$ shifts $k_T^-$, $k_L^-$ and $s_2^-$, and the component $\sigma$ shifts $f^-$ and $k_L^-$. Thus, there are three of the six source modes which are pure diffeomorphisms, matching the fact that there are only three independent vevs. The $\sigma$ diffeomorphism is not smooth at $r=0$, which will again be signalled by divergences in some of the functions near $r=0$. 

In the first part of our discussion below, this diffeomorphism freedom will not play a very important role, but when we come to solve the bulk equations numerically to investigate the stability of different boundary conditions, it will be convenient to work in terms of a set of diffeomorphism-invariant quantities. A useful set of diffeomorphism-invariant combinations of the functions are\footnote{Note that the last term in $g_3$ is not required for diffeomorphism-invariance, but is introduced for later convenience in writing the equations of motion.} 
\begin{equation}\label{def gs}
g_1 = j, \qquad  g_2 = k s_2 - \omega k_T,   \qquad g_3 = k ( z k_L - z k_T -  f) + 2 \omega s_1 
+ \frac{r^{2-2z}}{k}  (k s_2 - \omega k_T).
\end{equation}
In addition, the quantities 
\begin{align}\label{def fs}
	f_1 &= -k \omega \frac{r^{-2 z}  f}{4 z^2} +  \left(1 + \frac{\omega^2}{2 z^2 r^{2z}}\right) [ r^{2-2 z} \omega (\omega k_T - k s_2) +  s_1  ]  \\
	f_2 &= k^2  \frac{f}{4 r^2 z} +  k_T - \frac{k \omega}{2 z r^2} [  r^{2-2z} \omega (\omega k_T  -  k s_2) +  s_1]  \\
	f_3 &= k^2 \frac{f}{2 z} - \frac{k \omega}{z} [ r^{2-2 z} \omega (\omega k_T - k s_2) +  s_1]
\end{align}
transform as
\begin{equation}\label{diff transf sources}
 \delta f_1 = i k \phi^t, \qquad  \delta  f_2 = i k \phi^x, \qquad  \delta  f_3 = k^2 \sigma.
\end{equation}
Therefore, it is clear that $F_1 \equiv f_1'$, $F_2 \equiv f_2'$, $F_3 \equiv f_3'$ are also diffeomorphism invariant. Thus, if we make the change of variables from the original functions to $g_i, f_i$, the system of equations above will reduce to a system of six first order equations in the variables $g_i, F_i$. The source modes appearing in these functions are $s_\psi^-$ (which is unaffected by the diffeomorphisms) and the two diffeomorphism-invariant combinations 
\begin{equation}\label{a and b}
	a = \omega k_T^- - k s_2^-, \qquad b = z (k_L^-  - k_T^-) - f^- + 2 \omega k^{-1} s_1^-.
\end{equation}
Roughly speaking, the source $a$ is conjugate to $c_x$, while $b$ is conjugate to $c_t$. In the boundary condition where we fix the vevs, we will always work in a gauge where $s_1^- =0$, so that the source for the energy flux remains fixed as required (there is a diffeomorphism-invariant part in the source for the energy flux, but this is encoded in the vector part in our linearised analysis). In the stability analysis, we will also work in a gauge where the sources for the spatial stress tensor remain fixed. The choice of boundary conditions for the diffeomorphism-invariant modes will then correspond to the choice of either Dirichlet or Neumann boundary conditions for the momentum density and the energy density. 

\section{Boundary conditions}
\label{nbcs}

We now consider the inner product for the linearised perturbations reviewed in the previous section, and investigate the possible boundary conditions we can apply to them. There are two crucial constraints on the possible boundary conditions; they need to ensure conservation of the inner product by making the flux of the current through the boundary at infinity vanish, and the modes we allow to fluctuate need to have finite norm with respect to the inner product. We look for boundary conditions such that the fluctuating modes have finite norm with respect to the standard bulk inner product introduced in section \ref{IP}, without any explicit boundary terms. 

Before looking at the detailed calculations, we can discuss a few general expectations. We expect that normalizability in this standard inner product is connected to the absence of counterterms involving time derivatives of the boundary data in the boundary action. We would therefore expect the critical value of $z$ to be $z=d_s$, where the $r^{-2z}$ suppression associated with time derivatives becomes subleading compared to the $r^{-(z+d_s)}$ suppression of the terms in the action giving the finite vevs for operators relative to sources. So we expect that for $z >2$ (as we focus on $d_s = 2$) it will be possible to give alternative boundary conditions for some of the slow fall-off modes corresponding to sources. However, we will not be able to consider alternative boundary conditions for the modes corresponding to the sources for $\mathcal E^i$ and $\mathcal O_\psi$ for $z >2$, as these are irrelevant operators, so the corresponding modes grow in the UV, taking us outside of the scope of our linearised analysis. We will see that these are in any case always non-normalizable.

In the course of verifying the normalizability of the modes, we will also check the positivity of our inner product to the extent that we can; because we do not have a simple expression for the solution of the linearised equations for general momentum, we cannot fully verify this, but to the extent that we can calculate it we find that the inner product is indeed positive definite. 
 
\subsection{Flux through infinity}

To determine which boundary conditions lead to a conserved inner product, we need to consider the behaviour of the current in the asymptotic regime $r \to \infty$. This discussion can be carried out explicitly both for the zero-momentum and non-zero momentum modes, as it depends only on the form of the solution in the asymptotic regime. The flux is 
\begin{equation}
F = \int_{\partial M} \sqrt{\gamma} n^\mu j_\mu = \int dt d^{d_s} x r^{z+3} j_r. 
\end{equation}
we want to evaluate the flux between any two surfaces of constant time.

For the zero-momentum modes, the flux for the vector and tensor parts is 
\begin{equation}\label{flux vec k = 0}
    F = \frac{i}{2} \int_{\partial M} \left\{ \frac{2(z-1)}{z} r^{3z+1} [ v_{1i}^{(1)'} v_{1i}^{(2)*} - v_{1i}^{(1)} v_{1i}^{(2)*'}   ] +  r^{3+z} [ t_a^{(1)'} t_a^{(2)*} - t_a^{(1)} t_a^{(2)*'}    ] \right\},
\end{equation}
where we use $t_a$ to denote $t_o$, $t_d$, and we are allowing an arbitrary time dependence which is included in the functions. As we will see shortly, normalizability dictates that the only allowed boundary condition for the energy flux $\mathcal E_i$ are Dirichlet boundary conditions. Therefore we restrict ourselves to setting the linearised fluctuation $v_i^- = 0$ in our analysis of the flux. The leading behaviour of the modes is then obtained from (\ref{v3D k=0},\ref{v1 N}) giving 
\begin{equation}\label{v1 v3 k=0 near bndy explicit}
    v_{1i} =  v_i^+ r^{-3z} [1 + O(r^{-2z}) ] + \ldots,   \qquad  t_a = t_{a}^- \left[1 + \frac{\omega^2}{2z(2-z)} r^{-2z} \right] + t_{a}^+ r^{-(z+2)} + \ldots, 
\end{equation}
so we find that because we set $v_i^- = 0$, the flux only has a contribution from the tensor part, which is 
\begin{equation}\label{flux vec k = 0 2}
    F = \frac{i}{2} \int_{\partial M}  \left\{ (2+z) [ t^{-(1)}_{a} t^{+(2) \ast}_{a} - t^{-(2) \ast}_{a} t^{+(1)}_{a}  ] + \frac{i}{2(z-2)} (\omega_1^2 - \omega_2^2) t^{-(1)}_{a} t^{-(2) \ast}_{a} r^{2-z} \right\}. 
\end{equation}
The first term vanishes for real linear boundary conditions between $t_a^\pm$, but the second term produces a divergence for $z < 2$. We conclude that the flux vanishes for boundary conditions 
\begin{equation}\label{bcs vec sec k=0}
    v_i^- = 0, \quad t_{d}^- = \lambda_d t_{d}^+, \quad t_{o}^- = \lambda_o t_{o}^+,
\end{equation}
where the $\lambda$'s are arbitrary real numbers for $z>2$ and zero for $z<2$.

For the scalar part, we assume an $e^{i \omega t}$ dependence to simplify the expression by using the equations of motion. Then, the flux can be reduced to 
%
\begin{equation}\label{flux sc k = 0}
    F = \frac{i}{2} \int_{\partial M} e^{i(\omega_1-\omega_2)t} \left\{ r^{2-z} (\omega_1^2 - \omega_2^2) k_L^{(1)} k_L^{(2) \ast} +  \frac{r^{5+z}}{(z-1)} [ S^{(1)'} S^{(2) \ast} - S^{(1)} S^{(2)' \ast}    ] \right\}.
\end{equation}
The mode solutions in \eqref{kL soln} have the characteristic fall-offs at the boundary $r^{0}, r^{-(2 + z \pm \beta_z)/2}$. However the constant mode $k_L^{(0)}$ corresponds to a diffeomorphism which is not regular at $r=0$, and we will see in the next section that it has a divergent inner product as a result. Therefore we must set this mode to zero. For the modes associated with the scalar operator, the operator is irrelevant for $z >2$, and we will see that we must always take Dirichlet boundary conditions for this mode. Thus there is no choice of boundary conditions to make for the zero-momentum scalars. Plugging the allowed mode $k_L \sim r^{-(2 + z + \beta_z)/2}$ into \eqref{flux sc k = 0}, we can readily see that the flux vanishes. 

Turning to the non-zero momentum modes, the flux for the vector modes is given by 
\begin{align}
j_r =  \bigg[ \frac{(2-z)}{z} r^{2z-2} v_1^{(1)} v_1^{(2)'} + r^{2-2z} v_2^{(1)} v_2^{(2)'}  
&- v_1^{(1)} v_2^{(2)'}  - v_2^{(1)} v_1^{(2)'} \\
\nonumber
 &- v_3^{(1)} v_3^{(2)'} - (1 \leftrightarrow 2) \bigg],
\end{align}
where we write the expression for arbitrary dependence on the boundary coordinates, included in the functions. For $z <2$, setting all the source terms to zero will make the flux vanish. For $z > 2$, setting the source for the energy flux $\mathcal E_y$ to zero and plugging in the asymptotic solutions (\ref{v1a}-\ref{v3a}) for the other modes in a plane wave basis, we have that as $r \to \infty$, 
\begin{equation}
r^{z+3} j_r  = e^{i(\omega_1 - \omega_2) t+ i (k_1 - k_2) x} \left[ (v_2^{-(1)} k_2 c_y^{(2)} - v_2^{-(2)} k_1 c_y^{(1)}) - (v_3^{-(1)} \omega_2 c_y^{(2)} - v_3^{-(2)} \omega_1 c_y^{(1)}) \right], 
\end{equation}
so the flux will vanish and the inner product is conserved for general linear boundary conditions relating the sources $v_2^-, v_3^-$ to the vevs $kc_y$, $\omega c_y$, as expected.  In our stability analysis, we will focus on the simple cases of either Dirichlet or Neumann boundary conditions. In the Dirichlet case, we fix both $v_2^-$ and $v_3^-$; for the Neumann case, by contrast, we will fix $c_y = 0$. The fact that the flux vanishes implies that the diffeomorphism mode in the sources will be a null direction for the inner product (as we can give it a time dependence of compact support), and hence a pure gauge mode. 

For the scalar modes, the flux is given by
\begin{eqnarray}
j_r &=& e^{i(\omega_1 - \omega_2) t + i (k_1 - k_2) x} \left[k_L^{(1)} k_L^{(2)'}  - k_T^{(1)} k_T^{(2)'}  \right. \\ &&\nonumber +\left(\frac{(2-z)}{z} r^{2z-2} s_1^{(1)} s_1^{(2)'} + r^{2-2z} s_2^{(1)} s_2^{(2)'} - s_1^{(1)} s_2^{(2)'} - s_2^{(1)} s_1^{(2)'} \right) \\ &&\nonumber + 
\frac{2(z-1)}{z}  \left(j^{(1)} + \frac{1}{2} f^{(1)} \right) \left( j^{(2)'} + \frac{1}{2} f^{(2)'} \right) + k_L^{(1)} f^{(2)'} + f^{(1)} k_L^{(2)'} 
 \\ && \nonumber  + \frac{(z-1)}{r} j^{(1)} (2 k_L^{(2)} - f^{(2)}) - \frac{k}{2 z r^3} s_1^{(1)} (-2\omega_2 r^z (r^{1-z} k_L^{(2)})' + k r^{2z-1} (r^{2-2z} s_2^{(2)} - s_1^{(2)})') \\ && \nonumber \left. + \frac{\omega_2}{2z r^{1+2z}} \left(j^{(1)} + \frac{1}{2} f^{(1)} \right)  (-2\omega_2 r^z (r^{1-z} k_L^{(2)})' + k r^{2z-1} (r^{2-2z} s_2^{(2)} - s_1^{(2)})')- (1 \leftrightarrow 2) \right].
\end{eqnarray}

For $z <2$, setting all the source terms to zero will make the flux vanish.  For $z >2$, we want to see what boundary conditions are possible when we allow the source terms to be non-zero, although the source terms for $\mathcal E_x$ and $\mathcal O_\psi$ will remain zero, as these are irrelevant operators. There are potentially divergent terms in the flux involving products between the leading source terms and subleading terms given by derivatives of the leading source terms (as there were in the scalar field case). We do not calculate these explicitly, since we will see in the next section that the inner product is finite, so there cannot actually be any divergent terms in the flux - this would be incompatible with the finiteness of the inner product on arbitrary constant time surfaces. 

Plugging in the asymptotic solutions (\ref{fa}-\ref{s2a}), the finite part of the flux is given by
\begin{eqnarray}
\nonumber
r^{z+3} j_r  &=& e^{i(\omega_1 - \omega_2) t + i (k_1 - k_2) x} \bigg [  c_x^{(2)}( k_2 s_2^{-(1)} - \omega_2 k_T^{-(1)} ) - c_x^{(1)}( k_1 s_2^{-(2)} - \omega_1 k_T^{-(2)} )
\\
\nonumber
&+& \frac{k_2 c_t^{(2)}}{2} ( z k_L^{-(1)} - z k_T^{-(1)}  - f^{(1)} )  - \frac{k_1 c_t^{(1)}}{2} ( z k_L^{-(2)} - z k_T^{-(2)}  - f^{(2)} )  \bigg] \\
\nonumber 
&=& e^{i(\omega_1 - \omega_2) t + i (k_1 - k_2) x} \left[ \frac{1}{2} (k_T^{-(1)} \langle \Pi_{xx} - \Pi_{yy} \rangle^{(2)} - k_T^{-(2)} \langle \Pi_{xx} - \Pi_{yy} \rangle^{(1)})  \right. \\ && 
 + \frac{1}{2} (k_L^{-(1)} \langle \Pi_{xx} + \Pi_{yy} \rangle^{(2)} - k_L^{-(2)} \langle \Pi_{xx} + \Pi_{yy} \rangle^{(1)})\\ && \nonumber \left. + (s_2^{-(1)} \langle \mathcal P_x \rangle^{(2)} - s_2^{-(2)} \langle \mathcal P_x \rangle^{(1)}) - \frac{1}{2} (f^{(1)} \langle \mathcal E \rangle^{(2)}  -  f^{(2)}\langle \mathcal E \rangle^{(1)}) \right].
\end{eqnarray}
Thus, the flux vanishes for general linear boundary conditions relating the source and vev for these components.  Again, in the later stability analysis, we will focus on the simple cases of either Dirichlet or Neumann boundary conditions. For Dirichlet boundary conditions, both the diffeomorphism-invariants $a, b$ and the diffeomorphism modes must be fixed to make the flux vanish.  In the Neumann case, fixing $c_x, c_t$ makes the flux vanish, and as for the vector modes, the diffeomorphism mode in the sources will become a pure gauge mode.

\subsection{Inner product}

We now turn to considering the value of the inner product, which will determine when the slow fall-off modes corresponding to the boundary geometry are allowed to fluctuate. For the zero-momentum modes, we also compute the inner product explicitly to verify its positivity. 

\subsubsection{Zero momentum perturbations}
\label{ip0}

For the vector and tensor parts, the inner product  is
\begin{equation}\label{Omega vec k=0}
    ( \{h_1, a_1\} ,\{h_2, a_2\} ) = e^{i(\omega_1-\omega_2)t} Vol(x) \int_0^\infty dr {\cal F}_0(\omega_1, \omega_2; r),
\end{equation}
where $Vol(x)$ represents the spatial volume, and we recall that this inner product is independent of $t$ for boundary conditions such that the flux vanishes, and must therefore vanish for $\omega_1 \neq \omega_2$. The remaining integrand is 
\begin{equation}\label{cal F vec k=0}
    {\cal F}_0(\omega_1, \omega_2; r) = \frac{(\omega_1 + \omega_2)}{2} [ \alpha^2 r^{z-1} v^{(1)}_{1i} v^{(2)}_{1i} + r^{1-z} t^{(1)}_d t^{(2)}_d + r^{1-z} t^{(1)}_o t^{(2)}_o  ].
\end{equation}
We observe from \eqref{cal F vec k=0} that the different modes are orthogonal, as we expect, given that their equations of motion decouple. We see that the inner product is positive definite for $\omega>0$. The inner product for $v_{1i}$ seems to degenerate for $z=1$, where $\alpha=0$, but this is just an artifact of our choice of normalization $a_i \propto \alpha v_{1i}$, which was convenient for $z \neq 1$. For $z=1$, the massive vector mode $a_i$ still has a non-zero inner product. 

We can see from the first term that if we let the constant mode $v_i^-$ in $v_{1i}$ fluctuate, we would have a divergent inner product for any $z$. Thus, this mode is always non-normalizable. For simplicity we always take the boundary condition to be $v_i^-=0$.\footnote{It was shown in \cite{Ross:2011gu} that for $z < 2$  an asymptotic expansion still exists for a Dirichlet boundary condition fixing $v_i^-$ to some non-zero value. For $z >2$ however the extension to consider a non-zero source would be even more challenging than in the case of the scalar field.} For the tensor modes, we see that $t_o^-$, $t_d^-$ are non-normalizable for $z < 2$, but become normalizable for $z >2$. Thus we need to take Dirichlet boundary conditions for $z <2$ but we can take the general linear boundary conditions \eqref{bcs vec sec k=0} which conserve the inner product for $z >2$. 

Since we have closed form solutions for the bulk functions, we can evaluate the integrals to obtain the inner product explicitly. The integrals which appear in \eqref{cal F vec k=0} correspond precisely to the Sturm-Liouville products that follow from the respective decoupled differential equations. Thus, we write
\begin{equation}\label{cal F vec k = 0 SL}
    {\cal F}^{(v)}_0(\omega_1, \omega_2; r) = \frac{(\omega_1 + \omega_2)}{2} ( \alpha^2 \langle v^{(1)}_{1i} ,  v^{(2)}_{1i}\rangle +  \langle t^{(1)}_d ,  t^{(2)}_d\rangle + \langle t^{(1)}_o ,  t^{(2)}_o\rangle ),
\end{equation}
where the angle brackets denote the Sturm-Liouville products. As in the case of the scalar field, each Sturm-Liouville product can be rewritten as a total derivative of the Wronskian, converting the inner product to boundary contributions at $r = \infty$ and $r=0$, see e.g. \cite{Andrade:2011dg} for details. Setting $v_i^-=0$, the contribution from $v_{1i}$ gets only a contribution from the boundary term at $r=0$, which can be explicitly evaluated, giving
\begin{equation}\label{SL v1 result}
    \alpha^2 \frac{(\omega_1 + \omega_2)}{2} \langle v^{(1)}_{1i} ,  v^{(2)}_{1i}\rangle = \delta(\omega_1 - \omega_2) \frac{9 \pi z^3 \alpha ^2}{2 \omega ^3} |v_i^+ |^2.
\end{equation}
For the tensor modes, the linear boundary condition \eqref{bcs vec sec k=0} implies that the contribution from $r = \infty$ vanishes, and the term at $r=0$ gives 
\begin{equation}\label{SL v3 result}
    \frac{(\omega_1 + \omega_2)}{2}  \langle t^{(1)}_d ,  t^{(2)}_d \rangle = \delta(\omega_1 - \omega_2) | \lambda_d \kappa_{\omega,-a} + e^{i \pi a} \kappa_{\omega,a} |^2 |t_{d}^+|^2,
\end{equation}
where $\kappa_{\omega,a} = (2z)^2 \Gamma(1+a) \omega^{-a}$, and $\lambda_d$ is the parameter in the boundary condition \eqref{bcs vec sec k=0}, so $\lambda_d = 0$ for $z <2$. Similarly for $t_o$. The key point here is that these contributions to the inner product are manifestly positive definite. 

The inner product for zero-momentum scalar perturbations can be written as
\begin{equation}\label{Omega s k=0}
    ( \{h_1, a_1\} ,\{h_2, a_2\} ) = e^{i(\omega_1-\omega_2)t} Vol(x) \int_0^\infty dr {\cal G}^{(s)}_0(\omega_1, \omega_2; r),
\end{equation}
where
\begin{eqnarray}\label{cal G s k=0}
    {\cal G}^{(s)}_0(\omega_1, \omega_2; r) &=& \frac{(\omega_1 + \omega_2)}{2}r^{1-z} [\alpha^2  p^{(1)} p^{(2)} - k_L^{(1)} k_L^{(2)} ] \\
\nonumber
    &+& r (z-1) p^{(1)} \left[ j^{(2)} + k_L^{(2)} + \frac{r}{z} \left( j^{(2)'} + \frac{1}{2} F^{(2)} \right) \right] \\
\nonumber
    &+& r (z-1) p^{(2)} \left[ j^{(1)} + k_L^{(1)} + \frac{r}{z} \left( j^{(1)'} + \frac{1}{2} F^{(1)} \right) \right].
\end{eqnarray}

Using \eqref{alg F}, \eqref{alg p}, \eqref{alg j} and \eqref{eq S}, we can rewrite \eqref{cal G s k=0} as
\begin{equation}\label{cal G s k=0 kL}
    {\cal G}^{(s)}_0(\omega_1, \omega_2; r) = \frac{(\omega_1 + \omega_2)}{2}\left[  - \left( r^{2-z} k_L^{(1)} k_L^{(2)} \right)^\prime + \frac{r^{3-z}}{(z-1)} S^{(1)} S^{(2)}  \right].
\end{equation}

Note that the integral of the second term in \eqref{cal G s k=0 kL} is the Sturm-Liouville product for $S$ that follows from \eqref{eq S}. Therefore, the integral of \eqref{cal G s k=0 kL} can be written as
\begin{equation}\label{cal G s k=0 kL 2}
   \int_{0}^\infty {\cal G}^{(s)}_0 dr = \frac{(\omega_1 + \omega_2)}{2}\left[-r^{2-z} k_L^{(1)} k_L^{(2)}
   + \frac{r^{5+z}}{(z-1)} \frac{[ S^{(1)} S^{(2)' \ast} -  S^{(2)\ast} S^{(1)'}]}{(\omega_1^2 - \omega_2^2)} \right] \bigg|^{\infty}_{r=0}
\end{equation}
Note that the diffeomorphism mode $k_L^{-}$ enters only through a boundary contribution, as expected. However, for $z >2$ the boundary contribution at $r=0$ diverges. As discussed earlier, this is due to this not being a regular diffeomorphism at $r=0$; if we insist on working in radial gauge this mode must therefore be discarded.

If we let $s^-_\psi$ fluctuate, the first term on the right in \eqref{cal G s k=0 kL 2} will behave as $r^{-(2z - \beta_z)}$ as $r \to \infty$. Since $2z-\beta_z \leq 0$ for all $z$, this implies that $s^-_\psi$ is always non-normalizable. Thus we must take Dirichlet boundary conditions fixing $s^-_\psi$ and allowing $s^+_\psi$ to fluctuate. For simplicity we consider only $s^-_\psi = 0$.\footnote{Noting again that for $z >2$ this is the source for an irrelevant operator so it would be challenging to extend this to more general boundary conditions.}  

Using the explicit solution \eqref{kL soln}, the only non-zero contribution then comes from the second term in \eqref{cal G s k=0 kL 2} evaluated at $r=0$. Evaluating the Wronskian in \eqref{cal G s k=0 kL 2} explicitly, we find
\begin{equation}\label{ip kL prev}
   \int_{0}^\infty {\cal G}^{(s)}_0 dr =  |C_\omega|^2 \omega^{-\beta_z/z} |s^+_\psi|^2,
\end{equation}
where $C_\omega = (2z)^{\beta_z/z} \Gamma[1+\beta_z/(2z)] (2+z +\beta_z)/2$. 

Thus, we have seen explicitly that for these zero momentum perturbations, where we can calculate the inner product explicitly, it is manifestly positive definite.

\subsubsection{General perturbations} 

The inner product in the scalar sector for general $(\omega,k)$ reads
\begin{equation}\label{Omega s}
    ( \{h_1, a_1\} ,\{h_2, a_2\}) = e^{i(\omega_1-\omega_2)t} \delta(k_1-k_2) \int_0^\infty dr \left[ {\cal F}^{(s)}_0 + {\cal G}^{(s)}_0 + k_1 {\cal K}^{(s)} \right],
\end{equation}
\noindent where
\begin{equation}\label{cal F s}
    {\cal F}^{(s)}_0 = \frac{(\omega_1 + \omega_2)}{2} [ \alpha^2 r^{z-1} s^{(1)}_1 s^{(2)}_1 + r^{1-z} k_T^{(1)} k_T^{(2)} ],
\end{equation}
\begin{eqnarray}\label{cal G s}
    {\cal G}^{(s)}_0 &=& \frac{(\omega_1 + \omega_2)}{2}r^{1-z} [\alpha^2  p^{(1)} p^{(2)} - k_L^{(1)} k_L^{(2)} ] \\
\nonumber
    &+& r (z-1) p^{(1)} \left[ j^{(2)} + k_L^{(2)} + \frac{r}{z} \left( j^{(2)'} + \frac{1}{2} f^{(2)'} \right) \right] \\
\nonumber
    &+& r (z-1) p^{(2)} \left[ j^{(1)} + k_L^{(1)} + \frac{r}{z} \left( j^{(1)'} + \frac{1}{2} f^{(1)'} \right) \right],
\end{eqnarray}
and
\begin{eqnarray}\label{cal K s}
    {\cal K}^{(s)} &=& r^{z-1} \frac{(1-z)}{z} \left[ s^{(2)}_1  \left( j^{(1)} + \frac{f^{(1)}}{2} \right)  + s^{(1)}_1 \left( j^{(2)} + \frac{f^{(2)}}{2} \right) \right]  \\
    &+&  \frac{r^{z-1}}{2} [ s^{(1)}_1 ( k^{(2)}_T - k^{(2)}_L ) + s^{(2)}_1 ( k^{(1)}_T - k^{(1)}_L ) ] \\
    &+& \frac{r^{1-z}}{2} [ s^{(1)}_2 ( k^{(2)}_L - k^{(2)}_T ) + s^{(2)}_2 ( k^{(1)}_L - k^{(1)}_T ) ].
\end{eqnarray}
The function $p$ is determined algebraically as
\begin{equation}\label{p alg}
    p = \frac{r^{-z}}{4(z-1)} \left \{  2 \omega [ (z-1) k_L - r k'_L ] + k [ r s'_2 - 2(z-1) s_2 - r^{2z-1} s'_1 ]  \right\}.
\end{equation}

Since we only know the form of the mode solutions in the asymptotic region for these general perturbations, we can only verify the finiteness of the contribution to the inner product from the region $r \to \infty$. However, this is where we would expect to find divergences from allowing the slow fall-off modes to fluctuate. The sources for the stress tensor complex make $r$-independent contributions to the corresponding functions, so it is easy to see that most of them make a finite contribution to the inner product for $z >2$, as expected. An exception is the source $s_1^-$ for ${\cal E}_x$: from the term $r^{z-1} s_1^{(1)} s_1^{(2)}$ in \eqref{cal F s}, we conclude that this is non-normalizable for all values of $z$. Also, the source for the scalar operator $\mathcal O_\psi$ makes contributions to the functions at order $r^{-(z+2+\beta_z)/2}$, so this is also non-normalizable for all $z$, just as in the zero-momentum sector. 

There is also an exception in the other direction:  the $f^-$ source term in $f$ enters the inner product only through the combination $r^{z-1} s_1 f$ in \eqref{cal K s}. Since the source term in $s_1$ must be set to zero, we are left with $s_1 \sim r^{-(z+2)}, r^{-3z}, r^{-(5z+2+\beta_z)/2} $, so that the constant term in $f$ is normalizable for all $z$.\footnote{This is somewhat surprising, and might seem to conflict with the AdS results, as our analysis indicates that this mode is normalizable even for $z=1$. However,  at $z=1$, because the vector field vanishes, there is a gauge freedom which we can use to set this mode to zero, so it is pure gauge there. Note also that when we take the $f^-$ source term non-zero for $z<2$, there is a potentially divergent $f^- s_\psi^+$ term in the flux, but its coefficient vanishes, so the flux vanishes for a real linear boundary condition between $f^-$ and $c_t$ also for $z <2$. One could see that this coefficient has to vanish by the same argument used before; since the inner product evaluated on any surface of constant $t$ is finite, there can be no divergent terms in the flux.} These results are summarized in table \ref{table norm scalar}.

\begin{table}

\begin{center}
\begin{tabular}{|c|c|c|c|}
  \hline
  Operator   & Source  & $O(r^\alpha)$ &  Normalizability  \\
  \hline \hline
  ${\cal E}$             &  $f^-$    &      $r^0$      &     all $z$ \\
  \hline
  $\Pi^i_i$              &   $k_L^-$           &      $r^0$      &     $z>2$            \\
  \hline
  $\Pi_{xx} - \Pi_{yy}$  &   $k_T^-$           &      $r^0$      &     $z>2$            \\
  \hline
  ${\cal E}_x$           &   $s_1^-$           &      $r^0$      &     No           \\
  \hline
  ${\cal P}_x$           &   $s_2^-$           &      $r^0$      &     $z>2$        \\
  \hline
  $O_\psi$               &  $s_\psi^-$       &      $r^{-(z+2-\beta_z)/2}$      &     No        \\
  \hline
\end{tabular}
\caption{Normalizability for Neumann boundary conditions in the scalar sector. We emphasize that the source for ${\cal E}$ is normalizable for all $z$
{\it provided} the source for ${\cal E}_x$ vanishes.}
\label{table norm scalar}

\end{center}

\end{table}

\noindent In the vector sector, we find
\begin{equation}\label{Omega vec}
    ( \{h_1, a_1\} ,\{h_2, a_2\} ) = e^{i(\omega_1-\omega_2)t} \delta(k_1-k_2) \int_0^\infty dr [{\cal F}^{(v)}_0 + k_1 {\cal K}^{(v)}  ],
\end{equation}
\noindent where
\begin{equation}\label{cal F vec}
    {\cal F}^{(v)}_0 = \frac{(\omega_1 + \omega_2)}{2} [ \alpha^2 r^{z-1} v^{(1)}_1 v^{(2)}_1 + r^{1-z} v^{(1)}_3 v^{(2)}_3 ],
\end{equation}
and
\begin{equation}\label{cal K vec}
    {\cal K}^{(v)} = \frac{r^{-(z+1)}}{2} [ v_3^{(1)} ( r^{2z} v^{(2)}_1 - r^2 v^{(2)}_2 ) + v^{(2)}_3 ( r^{2z} v^{(1)}_1 - r^2 v^{(1)}_2 )     ].
\end{equation}
The sources for ${\cal E}_y$, ${\cal P}_y$ and $\Pi_{xy}$ are given by the constant terms in $v_1$, $v_2$, $v_3$, respectively. It is easy to see that the source for ${\cal E}_y$ is non-normalizable for all $z$, whereas the sources for ${\cal P}_y$ and $\Pi_{xy}$ are normalizable for $z>2$, see table \ref{table norm vector}.
\begin{table}

\begin{center}
\begin{tabular}{|c|c|c|c|}
  \hline
  Operator   & Source  & $O(r^\alpha)$ &  Normalizability  \\
  \hline \hline
  ${\cal E}_y$      &  $v_1^-$    &      $r^0$      &     No \\
  \hline
  ${\cal P}_y$      &  $v_2^-$    &      $r^0$      &     $z>2$            \\
  \hline
  $\Pi_{xy}$        &  $v_3^-$    &      $r^0$      &     $z>2$            \\
  \hline
\end{tabular}
\caption{Normalizability for Neumann boundary conditions in the vector sector.}
\label{table norm vector}
\end{center}

\end{table}

For the non-zero momentum modes, we cannot evaluate the inner product explicitly. As noted earlier, the vector and scalar parts will be orthogonal. However, even in the simpler vector sector, the equations are coupled and there is a two-dimensional space of solutions for given boundary conditions, so we cannot say anything about the positivity of the inner product just by writing it in terms of the functions $v_1, v_2, v_3$: we would need to write it in terms of the mode solutions. Since we know the solution only in the asymptotic region it is difficult to make real progress. 


To summarise, we see that for Lifshitz spacetimes for large enough $z$, we have the freedom to impose alternative boundary conditions for some of the geometrical boundary data. That is, we believe we can construct boundary field theories without ghosts by taking the definition of asymptotically locally Lifshitz spacetimes in \cite{Ross:2011gu} and switching from fixing the boundary data to more general boundary conditions relating the source and vev modes. This is a substantial difference from the more familiar AdS case, where \cite{Compere:2008us} showed that such alternative boundary conditions for the metric introduce ghosts. It is not obvious that such alternative boundary conditions will have interesting applications in condensed matter physics, but as they correspond in a certain sense to a dynamical boundary geometry, we hope that their exploration may teach us about gravitational aspects of the duality. 

\section{Stability for Neumann boundary conditions}
\label{spec}

In \cite{Andrade:2012xy}, we found that a scalar on a Lifshitz background can have an instability for Neumann boundary conditions. This is a new phenomenon in the Lifshitz case; in AdS, the conformal symmetry implies there can be no exponentially growing modes. For Lifshitz, the symmetry of the theory dictates the dispersion relation $\omega = \alpha k^z$, and we found that for Neumann boundary conditions, there are modes where $\alpha$ has a positive imaginary part, leading to exponential growth in time. In this section, we want to look for similar instabilities for the cases where linearised metric fluctuations satisfy Neumann boundary conditions.  

For the metric perturbations, scale invariance dictates that in the pure Lifshitz spacetime, imposing regularity at the horizon will fix the frequency to be $\omega = \alpha k^z$ for some $\alpha$, just as in the scalar field case, but because we do not have explicit solutions for non-zero $k$, we cannot determine $\alpha$ analytically. It is however still relatively straightforward to do a numerical analysis of the spectrum of modes. 

The instability, if it appears, will extend to arbitrarily high momentum, so the instability is essentially a UV effect. We will take advantage of this to regulate the IR region of the spacetime with a simple hard wall cutoff at $r=r_0$ in our numerical analysis. This will modify $\omega(k)$, which will generally be of the form $\omega = \alpha(r_0 k) k^z$. Assuming this gives an isolated solution as we remove the IR cutoff, $\alpha(r_0 k)$ has a non-zero limit for large argument, and we can find this limit by examining the behaviour for large $k$ for a fixed value of $r_0$. That is, we can find instabilities of the full Lifshitz spacetime by looking for instabilities of the hard wall solution which have $\omega = \alpha k^z$ for large enough $k$. 

\subsection{Vector perturbations}

For the numerical analysis of the vector modes, it is convenient to simplify the equations by differentiating \eqref{vec eom3} and solving \eqref{vec eom1} for $v_2'$, to write a pair of coupled equations for $v_1$ and $V_3 = v_3'$, 
\begin{eqnarray}
\nonumber
\omega (r^2 v_1'' + (1+ 3 z) r v_1')  +  k z r V_3 &=& \omega \left(\frac{k^2}{r^2} - \frac{\omega^2}{r^{2 z}} \right) v_1, \\
\label{v1V3}
r^2 V_3'' + r (5 + 3z) V_3' + (2z^2 + 7 z + 3) V_3 + 2 (z-1) k \omega \frac{v_1}{r^3} &=& \left( \frac{k^2}{r^2}- \frac{\omega^2}{r^{2 z}} \right) V_3.
\end{eqnarray}
The asymptotics given in (\ref{v1a},\ref{v3a}) then imply
\begin{eqnarray}
\nonumber
	v_1  &=& v_1^- +  \frac{z k^2 \omega c_y}{2(z-1)(z+2)} r^{-(z+2)} +  \frac{k^2 v_2^{-}}{2z(2-z)} r^{-2z} + v_1^+ r^{-3z}, \\
\label{short asympt v1 V3} 	
	(k \omega)^{-1} \ V_3 &=& - \frac{ v_1^-}{z} r^{-3} +  \omega c_y r^{-(z+3)}  +\frac{v_2^-}{2-z} r^{-(2z+1)} + \frac{(1-z)}{z(2+z)} v_1^+ r^{-(3z+3)} .
\end{eqnarray}
In the IR, we introduce a hard wall cutoff which we take to be at $r=1$ without loss of generality. We choose to impose a Dirichlet boundary condition at this wall, fixing $V_3(1) =0$, $v_1(1)=0$. This boundary condition ensures the vanishing of the flux through $r=1$. We always restrict to the boundary condition $v_1^-=0$. We then want to determine the spectrum for the Neumann boundary condition $c_y = 0$, and for the conventional Dirichlet boundary condition $v_3^- = 0$ (we have fixed the diffeomorphism symmetry by dropping the constant mode in $v_2$, so this really corresponds to a boundary condition on the diffeomorphism-invariant combination $\omega v_3^- - k v_2^-$). Thus, the boundary conditions of interest are
\begin{equation} \begin{array}{cl} 
\textrm{Neumann:} & v_1(1) = 0, V_3(1)=0, v_1^-=0, c_y=0; \\
\textrm{Dirichlet:} & v_1(1) = 0, V_3(1) = 0, v_1^-=0, v_3^-=0. \end{array}
\end{equation}
As mentioned in section \ref{vp}, we interpret these as specifying either Neumann or Dirichlet boundary conditions for the momentum density $\mathcal P_y$. 

\begin{figure}[htb]
\center
\subfigure[][]{
\label{vec NM N z 5/2}
\includegraphics[width=0.40\linewidth]{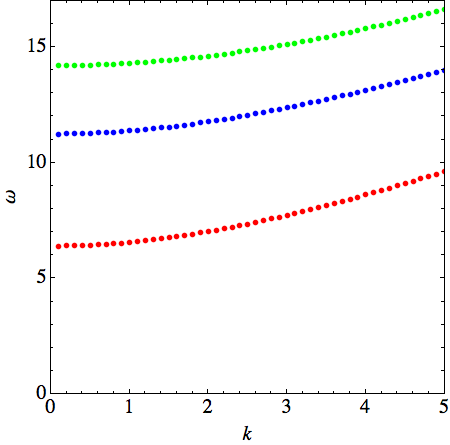}
}\qquad\qquad
\subfigure[][]{
\label{vec NM N z 7/2}
\includegraphics[width=0.40\linewidth]{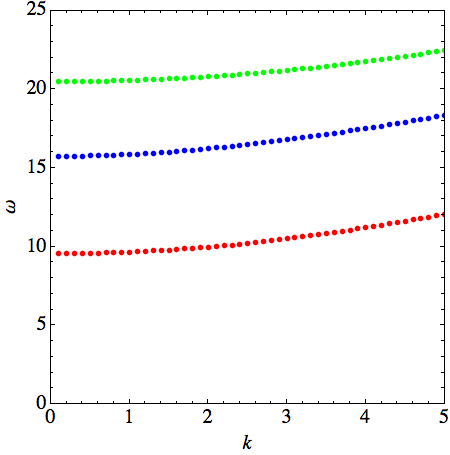}
}
\caption{Series of first vector normal modes in the theory with Neumann boundary conditions for $z = 5/2$ (left), $z = 7/2$ (right).}
\end{figure}

The most  convenient approach for numerical investigation is to discretize the eigenvalue problem using spectral methods. This is possible if the target functions are analytic, which can be achieved by a suitable choice of variables for half-integer $z$. For Neumann boundary conditions, if we define $r = y^{-2}$ and introduce the functions 
\begin{equation}
	x_1(y) = r^{2z} v_1, \qquad \qquad x_2(y) = r^{2z+1} V_3,
\end{equation}
then the Neumann boundary conditions become 
\begin{equation} 
 x_1(1) = 0, x_2(1)=0, x_1'(0)=0, x_2'(0)=0,
 \end{equation}
and the desired target functions are of the form $x_1 \sim y^0, y^{2z}$ and $x_2 \sim y^0, y^{2z+4}$. Carrying out this calculation for $z = 5/2, 7/2$, we find that there are no instabilities; the eigenvalue problem gives only real frequencies $\omega$. This result was cross-checked by a shooting method, working in a variable $u = 1/r$, numerically integrating from the wall to the boundary and reading off the coefficients at the boundary by fitting with the asymptotic expansions. The results for the eigenvalues obtained by the two methods agree well. In figures \ref{vec NM N z 5/2}, \ref{vec NM N z 7/2} we show the dispersion relation of the smallest 
vector normal modes for $z = 5/2, 7/2$ with Neumann boundary conditions. 

\begin{figure}[htb]
\center
\subfigure[][]{
\label{vec NM D z 5/2}
\includegraphics[width=0.40\linewidth]{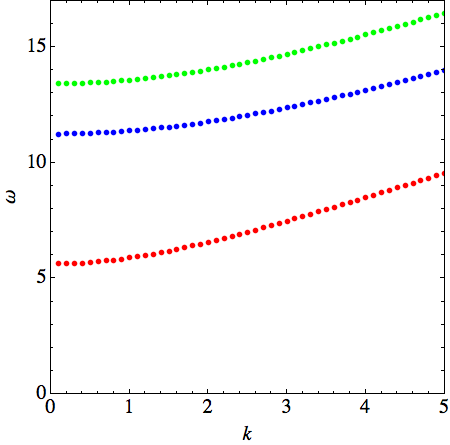}
}\qquad\qquad
\subfigure[][]{
\label{vec NM D z 7/2}
\includegraphics[width=0.40\linewidth]{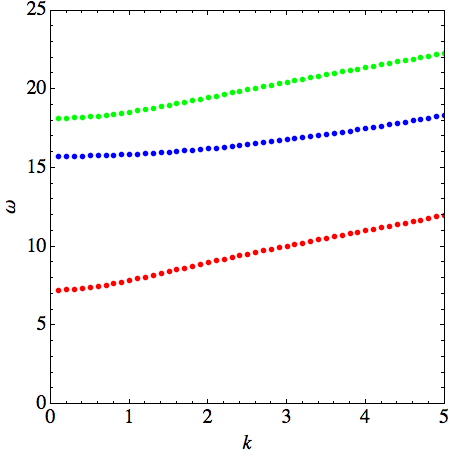}
}
\caption{Series of first vector normal modes in the theory with Dirichlet boundary conditions for $z = 5/2$ (left), $z = 7/2$ (right).}
\end{figure}

Similarly, for Dirichlet boundary conditions, if we define $y= r^{-1}$ and introduce the functions 
\begin{equation}
	x_1(y) = r^{z+2} v_1, \qquad \qquad x_2(y) = r^{z+3} V_3,
\end{equation}
then the Dirichlet boundary conditions become 
\begin{equation} 
 x_1(1) = 0, x_2(1)=0, x_1'(0)=0, x_2'(0)=0,
 \end{equation}
and the desired target functions are of the form $x_1 \sim y^0, y^{2z+2}$ and $x_2 \sim y^0, y^{2z}$. Carrying out this calculation for $z = 5/2, 7/2$, we again find that there are no instabilities; the eigenvalue problem gives only real frequencies $\omega$, see figures \ref{vec NM D z 5/2}, \ref{vec NM D z 7/2} for the first normal modes we obtain for Dirichlet boundary conditions. 

Thus, we see that at least for some values of $z$, both the Dirichlet and Neumann boundary conditions give rise to stable bulk theories. This is a proof of principle that it is possible to take these alternative boundary conditions for metric modes. It might be interesting in future to explore more thoroughly to verify if this stability depends on the value of $z$. 
 
\subsection{Scalar perturbations}

In the scalar case, in solving the bulk equations it is convenient to work in terms of the diffeomorphism-invariant functions $g_i, F_i$ for $i= 1,2,3$ introduced in (\ref{def gs},\ref{def fs}). An appropriate set of boundary conditions at the IR wall which make the flux vanish are $g_i = 0$, $f_i = 0$ at $r=1$, which leaves $F_i(1)$ as the free IR data. In the UV, we will always restrict to the boundary conditions $s_\psi^- = 0$. There is then a choice of boundary conditions for the four remaining pieces of diffeomorphism-invariant boundary data $a$, $b$, $c_t$, $c_x$.
We consider taking Dirichlet or Neumann boundary conditions for these. As mentioned in section \ref{sp}, we interpret this as taking either Dirichlet or Neumann boundary conditions for the momentum density and the energy density.  The source term for the energy flux must be set to zero, $s_1^- = 0$, but in the scalar sector this can always be achieved by suitably choosing the diffeomorphism, so it does not restrict the boundary conditions for the diffeomorphism-invariant modes.  The boundary conditions of interest are then 
\begin{equation} \begin{array}{lll}\label{UV bcs scalar sector}
\textrm{Dirichlet } \mathcal P_x, &\textrm{Dirichlet } \mathcal E: & g_i(1) = 0, s_\psi^-=0, a =0, b=0, \\
\textrm{Neumann } \mathcal P_x, &\textrm{Dirichlet } \mathcal E: & g_i(1) = 0, s_\psi^-=0, c_x=0, b=0, \\
\textrm{Dirichlet } \mathcal P_x, &\textrm{Neumann } \mathcal E: & g_i(1) = 0, s_\psi^-=0, a =0, c_t=0, \\
\textrm{Neumann } \mathcal P_x, &\textrm{Neumann } \mathcal E: & g_i(1) = 0, s_\psi^-=0, c_x =0, c_t=0,
\end{array}
\end{equation}
where $a, b$ are the diffeomorphism-invariant combinations of the sources introduced in  \eqref{a and b}. 

\begin{figure}[htb]
\center
\subfigure[][]{
\label{IR inst z 5/2}
\includegraphics[width=0.40\linewidth]{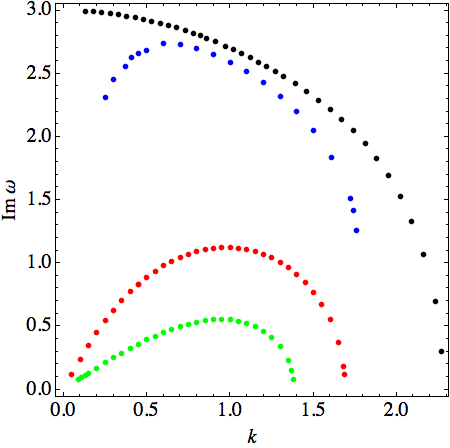}
}\qquad\qquad
\subfigure[][]{
\label{IR inst z 7/2}
\includegraphics[width=0.40\linewidth]{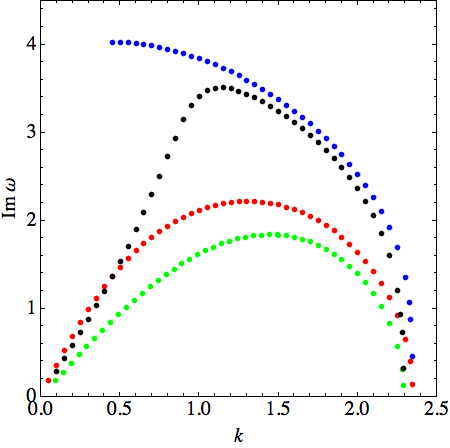}
}
\caption{Purely imaginary frequency modes for small $k$ with the boundary conditions $a=b=0$ (green); $b=c_x = 0$ (red); 
$c_t=c_x=0$ (blue); $a = c_t= 0$ (black). On the left we show $z=5/2$ while on the right we have $z=7/2$}
\end{figure}

\begin{figure}[htb]
\center
\subfigure[][]{
\label{NM NPx z 5/2}
\includegraphics[width=0.40\linewidth]{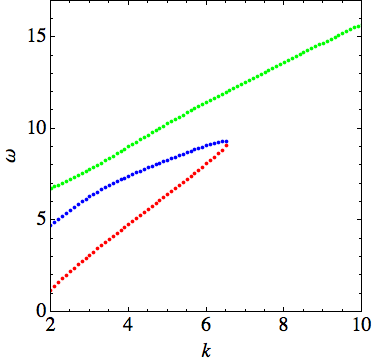}
}\qquad\qquad
\subfigure[][]{
\label{NM NPx z 7/2}
\includegraphics[width=0.40\linewidth]{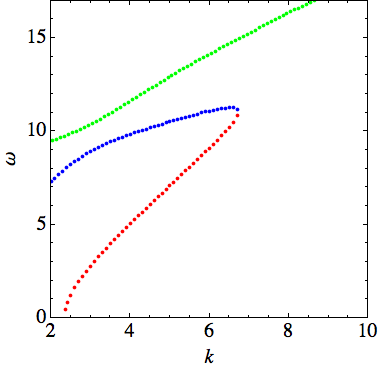}
}
\caption{First series of normal modes for $b = c_x = 0$ with $z = 5/2$ (left) and $z=7/2$ (right).}
\end{figure}

\begin{figure}[htb]
\center
\subfigure[][]{
\label{re cplex soln z 5/2}
\includegraphics[width=0.40\linewidth]{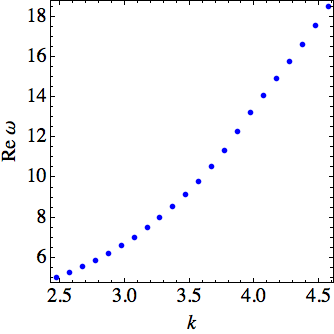}
}\qquad\qquad
\subfigure[][]{
\label{im cplex soln z 5/2}
\includegraphics[width=0.40\linewidth]{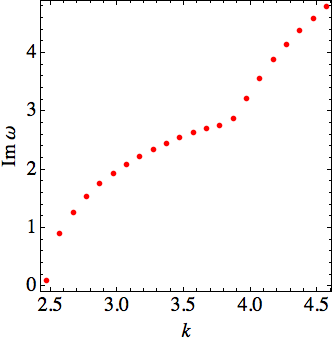}
}
\caption{Complex solutions that appear at large momentum for the boundary condition $a = c_t = 0$ for $z = 5/2$. We plot the momentum dependence of the real and imaginary  parts of the frequencies on the left and right figures, respectively.}
\end{figure}

\begin{figure}[h]
\begin{center}
\includegraphics[scale=0.40]{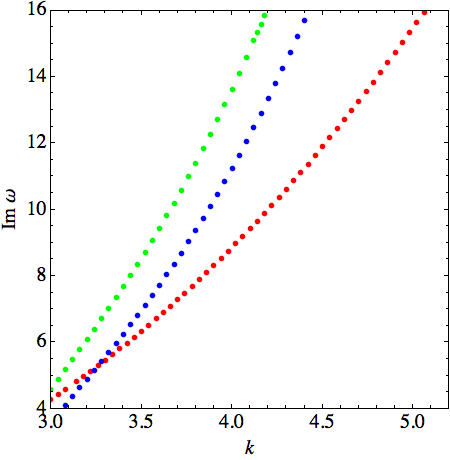}
\caption{Purely imaginary frequency solutions persist at large $k$ for $c_t = c_x = 0$ with $z = 5/2$ (red); $c_t = c_x = 0$ with $z = 7/2$ (blue); $c_t = a= 0$ with $z = 7/2$ (green). The dispersion relation is of the form $\omega \propto k^{z}$.  }
\label{UV inst}
\end{center}
\end{figure}

In this case we simply solve for the spectrum using a shooting procedure: we integrate outwards from the wall and inwards from the boundary and match the solutions at an intermediate point, which gives six conditions that we need to satisfy (recall that our system is first order, so we do not need to match the derivatives). Before presenting our numerical results, it is illustrative to perform a simple counting of degrees of freedom. As mentioned above, we have three independent pieces of boundary data in the IR, one of which can be fixed by linearity. For any of the choices \eqref{UV bcs scalar sector} in the UV, we have three degrees of freedom. Thus, for fixed $k$, we expect to find solutions for a discrete set of values of $\omega$, since there are six parameters to satisfy six matching conditions. 
We solve the matching equations at the intermediate point using a Newton-Raphson algorithm, taking random seeds with complex frequency. For concreteness, we focus on $z = 5/2$ and $z = 7/2$. We can summarize our findings as follows:

\begin{itemize}

\item In all the cases studied, we find IR instabilities corresponding to purely imaginary frequency modes that do not persist above some critical value of $k$, which we denote as $k_0$ (the precise value depends on the theory of interest). The momentum dependence of these modes can be found in figures \ref{IR inst z 5/2}, \ref{IR inst z 7/2}. It is worth mentioning that the analytic black hole solution of \cite{Balasubramanian:2009rx}, \cite{Giacomini:2012hg} exhibits a similar kind of IR instability, as noted in \cite{Andrade:2012xy}.

\item For Neumann boundary conditions for $\mathcal P_x$ but not for $\mathcal E$, the procedure converges to real frequencies $\omega$ for large enough $k$. Thus, there is no sign of a UV instability for these boundary conditions. In figures 
\ref{NM NPx z 5/2}, \ref{NM NPx z 7/2} we show our results for the $k$ dependence of the frequency of the first normal modes for $z=5/2$ and $z = 7/2$. 

\item For Neumann boundary conditions for $\mathcal E$ and not for $\mathcal P_x$, we find instabilities for $k > k_0$, which we thus interpret as a UV issue. These correspond to complex frequency solutions for $z = 5/2$, see figures \ref{re cplex soln z 5/2}, \ref{im cplex soln z 5/2} and purely imaginary frequency solutions for $z = 7/2$, see figure \ref{UV inst}.  In the former case, our algorithm suffers from large numerical uncertainty above $k \geq 4.5$, which has prevented us from explicitly verifying the expected dispersion relation, $\omega \propto k^z$. 
For $z = 7/2$, on the other hand,  we can verify that the purely imaginary frequencies behave approximately as $\omega = i \beta(z) k^z$ with  
\begin{equation}
	\beta_{c_t = a = 0}(7/2) =0.11.
\end{equation}

\item For Neumann boundary conditions for $\mathcal E$ and $\mathcal P_x$, we find a UV instability for both $z= 5/2$ and $z = 7/2$, corresponding to purely imaginary frequency solutions approaching $\omega = i \beta(z) k^z$ for large $k$, see figure \ref{UV inst}. The values of $\beta(z)$ are given by
\begin{equation}
	\beta_{c_t = c_x = 0}(5/2) = 0.27, \qquad \beta_{c_t = c_x = 0}(7/2) = 0.08.
\end{equation}

\item Finally, for Dirichlet boundary conditions for both $\mathcal P_x$ and $\mathcal E$, we do not find instabilities for $k >k_0$.

\end{itemize}

 It is very interesting that we find no UV instability in both the vector and scalar parts for Neumann boundary conditions for the momentum density $\mathcal P$, but that we did find one when we considered Neumann boundary conditions for the energy density $\mathcal E$. This may well be correlated with the fact that the momentum density is a relevant operator in a Lifshitz theory, so that its source has a positive conformal dimension, while the energy density remains a marginal operator, so that its source has dimension zero. We do not know from the field theory point of view what the lower bound on the dimensions of operators is, but it is not surprising to find that dynamical operators of vanishing conformal dimension appear to be excluded. It is also not unexpected that alternative boundary conditions for $\mathcal P$ are allowed for some values of $z$, since we know that at $z \geq 4$ the source and vev modes cross over and the boundary condition on the bulk field fixing the mode which falls off more slowly for $z \geq 4$ corresponds to fixing the vev rather than the source of $\mathcal P$, as we discuss in appendix \ref{z4}.


\section*{Acknowledgements}

We are grateful for useful conversations with Diego Hofman, Shamit Kachru, Don Marolf, Mukund Rangamani, Eva Silverstein, David Tong, Helvi Witek, and Ben Withers. This work was supported by STFC by the National Science Foundation under Grant No. NSF PHY11-25915 and Grant No PHY08-55415, and by funds from the University of California.

\appendix

\section{Asymptotic symmetries}
\label{symm}

As we noted in section \ref{bdiff}, there are bulk diffeomorphisms which act non-trivially on the boundary data. The subset of these diffeomorphisms which preserve a given choice of boundary conditions will define the asymptotic isometries of our spacetime. Asymptotic isometries have not yet been explored in detail for the asymptotically locally Lifshitz boundary conditions of \cite{Ross:2011gu}, although asymptotic symmetry groups were discussed in a related context in \cite{Compere:2009qm}. Therefore, in this appendix, we give a general abstract discussion of the possible asymptotic isometry groups for our different boundary conditions. This discussion will consider the possible asymptotic symmetries for any spacetime satisfying our boundary conditions in the UV, and not just for linearized perturbations around pure Lifshitz. 

The bulk diffeomorphisms of interest are asymptotically of the form \eqref{diffeo}, corresponding to a boundary diffeomorphism $\phi^\mu$ and a Weyl scaling $\sigma$ acting on the boundary geometry.  The timelike basis vector $e^{(0)}$ is also distinguished by the asymptotic boundary conditions, so the local Lorentz symmetry acting on the basis vectors gets restricted to the rotational symmetry $M^I_{\ J}$ acting on the spacelike vectors. The action of these transformations on the boundary data is 
\begin{equation}
\delta \hat e^{(0)}_\alpha = z \sigma \hat e^{(0)}_\alpha + \mathcal{L}_\phi \hat e^{(0)}_\alpha, \quad \delta \hat e^{(I)}_\alpha =  \sigma \hat e^{(I)}_\alpha + \mathcal{L}_\phi \hat e^{(I)}_\alpha
+ M^I_{\ J} \hat e ^{(J)}_\alpha, \end{equation}
where $\mathcal L_\phi$ is the Lie derivative along $\phi^\alpha$. These represent a combination of an anisotropic Weyl transformation given by $\sigma(x^\mu)$, a boundary diffeomorphism, and a local frame rotation of the spacelike frame vectors. 

In \cite{Ross:2011gu} two possible notions  of asymptotically locally Lifshitz spacetimes were proposed: either leaving $\hat e^{(0)}$ as arbitrary boundary data, or restricting to  $\hat e^{(0)}_i = 0$, to ensure that the boundary geometry had a global foliation by surfaces of constant $t$. The latter restriction would further restrict the allowed transformations:
\begin{equation}
\delta \hat e^{(0)}_i = \mathcal L_\phi \hat e^{(0)}_i = \hat e^{(0)}_t \partial_i \phi^t = 0, 
\end{equation}
so we are restricted to boundary diffeomorphisms where $t'(t)$ only, as expected. Geometrically,  requiring that the boundary has a foliation by surfaces of constant time restricts us to consider foliation-preserving diffeomorphisms. 

The asymptotic symmetries are the subset of these transformations which preserve a specific choice of boundary data. When we consider the Dirichlet boundary conditions, the asymptotic symmetries are the transformations that preserve the given $\hat{e}^{(0)}$, $\hat{e}^{(I)}$. The choice of a non-degenerate set of spacelike frame fields $\hat e^{(I)}$ leaves no freedom to do frame rotations as an independent symmetry, but we may still have a compensating frame rotation for another transformation. Any transformation that leaves the ``spacelike metric"  $g_{\alpha\beta} = \hat e^{(I)}_\alpha \hat e_{(I) \beta}$ invariant can thus be promoted to a symmetry that leaves $\hat e^{(I)}_\alpha$ invariant by considering an appropriate frame rotation. Thus we can characterise the transformations that preserve the given boundary data as the boundary diffeomorphisms $\phi$ satisfying
\begin{equation} \label{CKV}
\mathcal{L}_\phi \hat e^{(0)}  = -z \sigma(x) \hat e^{(0)}, \quad \mathcal{L}_\phi \hat g_{\alpha \beta} = - 2 \sigma(x) \hat g_{\alpha \beta}
\end{equation}
for some Weyl transformation $\sigma(x^\mu)$. These are the natural anisotropic analogues of conformal Killing vectors. This is what we would expect as the natural analogue of the AdS result. In particular, choosing the pure Lifshitz boundary condition 
\begin{equation}
\hat e^{(0)} = dt, \quad \hat e^{(I)} = dx^i,
\end{equation}
these symmetries will be the Lifshitz symmetries, $H, P_i, M_{ij}, D$.\footnote{Note that the asymptotic symmetries are just the isometries in any dimension; this differs from the result in \cite{Compere:2009qm}, where it was argued that there will be an infinite-dimensional group of asymptotic symmetries for a three-dimensional bulk. This is presumably due to differences in the definition of the boundary conditions.}

We can determine the transformation of the stress tensor under the bulk diffeomorphism using the invariance of the action. This gives
\begin{equation}
\delta T^\alpha_{(0)} = - (2z + d_s) \delta \sigma T^\alpha_{(0)} + \mathcal L_\phi T^\alpha_{(0)}, \quad \delta T^\alpha_{(I)} = - (z + d_s +1) \delta \sigma T^\alpha_{(I)} + \mathcal L_\phi T^\alpha_{(I)} + M^J_{\ \ I} T^\alpha_{(J)}.
\end{equation}
The extra factor of $z+d_s$ in the Weyl scaling of $T^\alpha_{(A)}$ comes from the measure factor in the action. For $z > 2$, if we consider for example the maximally Neumann boundary conditions, with Neumann for all components except for the energy flux, 
\begin{equation} \label{Nbc}
\hat e^{(0)}_i = 0, \quad T^t_{(0)} = T^\alpha_{(I)} = 0,
\end{equation}
this is invariant under the anisotropic Weyl rescaling, and under all foliation-preserving boundary diffeomorphisms $t'(t), x^{i'}(t, x^i)$. We can also consider various mixed Neumann and Dirichlet boundary conditions, as for example the theory with Neumann boundary conditions for the momentum density and Dirichlet boundary conditions for the energy density which we found to be stable.

For the Dirichlet boundary conditions, the conformal Killing transformations are global transformations from the boundary point of view, and have non-zero associated conserved charges. For Neumann boundary conditions, the unfixed diffeomorphisms which are local symmetries will also be gauge transformations from the point of view of the boundary theory: if we  evaluate the associated conserved charge on a slice $\Sigma$ on the boundary where the transformation vanishes, it will trivially vanish. Similarly they are null directions for the inner product.

The boundary conditions \eqref{Nbc} have the same symmetries as a Horava-Lifshitz gravity theory \cite{Horava:2009uw}, so one might wonder about the relation. In \cite{Griffin:2011xs}, the boundary counter terms in the Dirichlet theory for $z=2$ were found to reproduce the action of a Horava-Lifshitz theory in $d=3$ spacetime dimensions. However, this is outside of the regime $z >2$ where we can apply Neumann boundary conditions. This is not a coincidence; the action for the Horava-Lifshitz theory naturally contains a dynamical term involving $(\partial_t \hat e)^2$; as we argued before it is precisely when such terms are not required as counter terms that we can introduce generalised boundary conditions for the metric while using the standard inner product. This implies more generally that as in the AdS case, it is not possible to make the boundary geometry truly dynamical by adding a explicit boundary action for the boundary geometry; the kinetic term in such a boundary action is an irrelevant deformation of the boundary theory for $z >2$. 

\section{Boundary conditions for $z > 4$ and asymptotic expansion}
\label{z4}

In this appendix, we discuss the asymptotic expansion for $z >4$, where the mode corresponding to the expectation value of the momentum density dominates over the mode corresponding to the source for the momentum density at large $r$.\footnote{This issue was briefly mentioned in \cite{Ross:2011gu} but was not addressed in detail there.}  This discussion is somewhat tangential to the main purpose of our paper, and can to some extent be read independently of the main text. 

For $z >4$, we need to modify the basic definition of asymptotically locally Lifshitz boundary conditions. As argued in \cite{Ross:2011gu}, we can consider a modified Dirichlet boundary condition, requiring $\hat e^{(0)}_i = 0$ and requiring that $\hat e^{(0)}_t$, $\hat e^{(I)}_j$ and $\tilde e^{(I)}_t =
r^{4-z} \hat e^{(I)}_t$ are finite as $r \to \infty$. The connection to the rest of the paper is that this modified Dirichlet boundary condition corresponds in the field theory to fixing $\mathcal P_i$ and letting the corresponding source fluctuate. However, from the spacetime point of view this is a Dirichlet boundary condition, fixing the leading part of the frame field in the asymptotic regime, so the relevant analysis is closer to that in  \cite{Ross:2011gu} than to the main body of the paper.

For this new boundary condition, the first question we need to ask is whether the asymptotic expansion remains valid; that is, for arbitrary values of $\hat e^{(0)}_t$, $\hat e^{(I)}_j$ and $\tilde e^{(I)}_t$, can we build a solution of the bulk equations of motion by adding subleading terms in a large $r$ expansion? The leading order fields have
\begin{equation}
\hat e^{(0)}_t, \hat e^{(I)}_j \sim r^0, \quad \hat e^{(0)}_i = 0, \quad \hat e^{(I)}_t \sim r^{z-4},
\end{equation}
so the inverse frame fields have
\begin{equation}
\hat e_{(0)}^t, \hat e_{(I)}^j \sim r^0, \quad \hat e_{(0)}^i \sim r^{z-4}, \quad \hat e_{(I)}^t =0.
\end{equation}
The leading behaviour of the frame extrinsic curvature is then
\begin{equation}
K_{00}, K_{IJ} \sim r^0, \quad K_{0I} = 0, \quad K_{I0} \sim r^{-3}, 
\end{equation}
and the leading behaviour of the Ricci rotation coefficients is 
\begin{equation}
\Omega_{0I}^{\ \ 0}, \Omega_{IJ}^{ \ \ \ K} \sim r^0, \quad \Omega_{IJ}^{\ \ \ 0} =0, \quad \Omega_{0I}^{\ \ J} \sim r^{-4} + r^{-z}. 
\end{equation}
The behaviour of the derivatives with frame indices is also affected, $\partial_0 \sim r^{-z} \partial_t + r^{-4} \partial_i$, $\partial_I \sim r^{-1} \partial_i$. The fact that all these powers are still negative indicates that we will be able to construct an asymptotic expansion.

The source terms in the radial expansion of the equations of motion involve $\nabla^A F_{AB}$ and $R_{AB}$ in addition to the frame extrinsic curvature; the above behaviour gives 
\begin{equation}
\nabla^A F_{A0} \sim r^{-2}, \quad \nabla^A F_{AI} \sim r^{-(1+z)}, r^{-5}, 
\end{equation}
and
\begin{equation}
R_{00}, R_{IJ}  \sim r^{-2}, r^{-2z}, r^{-z-4}, r^{-8}, \quad R_{0I} \sim r^{-(1+z)}, r^{-5}, \quad R_{I0} = 0.
\end{equation}
to cancel the source terms, we will then need the first subleading terms in the frame fields to appear at 
\begin{equation}
\hat e^{(0)}_t \sim r^{-2}, r^{-2z}, r^{-z-4}, r^{-8}, \quad \hat e^{(0)}_i \sim r^{-z-4}, 
\end{equation}
\begin{equation}
\hat e^{(I)}_t \sim r^{-2}, r^{z-6}, \quad \hat e^{(I)}_j \sim r^{-2}, r^{2-z}.  
\end{equation}
These are all subleading compared to the leading terms; the main novelty relative to the usual asymptotically locally Lifshitz boundary conditions is that odd powers of $z$ now appear. So in general we should be able to construct an asymptotic expansion in powers of $r$ involving an expansion in $r^{-2m - 2zn - (z-4)p}$ relative to the leading terms.

In this radial expansion, the subleading terms are explicitly determined as derivatives of the leading boundary data, so any divergences involving these terms can be cancelled by local counter terms. To explicitly evaluate the counterterms, one could carry out the holographic renormalization as in \cite{Ross:2011gu}. 

Following the analysis in this paper, we have seen that both the source mode $s_2^-$ and the vev mode $c_x$ are normalizable for any $z >2$. So both the above modified Dirichlet boundary condition and the alternative boundary condition where we continue to fix the now subleading mode $s_2^-$ for $z >4$ are acceptable from the point of view of the inner product in the linearised theory. 
Further exploration of the theory for $z > 4$, and the investigation of the renormalization group flow sourced by turning on the deformation by $\mathcal P_i \mathcal P^i$, which is relevant for $z>4$, is left for future work. 

\bibliographystyle{utphys}
\bibliography{lifshitz}

\end{document}